\documentclass[10pt,aps,prb,amsmath,amssymb,twocolumn,letterpaper,
         citeautoscript,showpacs,raggedbottom,balancelastpage]{revtex4-1}
\usepackage[usenames,dvipsnames]{color}
\usepackage[pdftex,
  breaklinks,             %
  bookmarks = false, %
  pdfpagemode   = UseNone,%
  pdfstartview 	= FitH,		%
  pdfstartpage  = 1,      %
  colorlinks    = true,   %
]{hyperref}
\hypersetup{
    linkcolor=RubineRed,          
    citecolor=ForestGreen,        
    filecolor=Mulberry,      
    urlcolor=RoyalBlue           
}
\usepackage{graphicx}
\usepackage[protrusion=true,expansion=true,final]{microtype}
\usepackage{tikz}
\usepackage{longtable}



\begin{document}
%

\title{Interplay of octahedral rotations and breathing 
	distortions in charge ordering perovskite oxides}

\author{Prasanna V.\ Balachandran }
	\affiliation{Department of Materials Science \& Engineering,\!
	Drexel University,\! Philadelphia,\! PA 19104,\! USA}%
\author{James M.\ Rondinelli}
  \email{jrondinelli@coe.drexel.edu}
	\affiliation{Department of Materials Science \& Engineering,\!
	Drexel University,\! Philadelphia,\! PA 19104,\! USA}%
\date{\today}
\begin{abstract}
We investigate the structure--property relationships in 
$AB$O$_3$ perovskites  exhibiting octahedral rotations and cooperative octahedral 
breathing distortions (CBD) using group theoretical methods. 
Rotations of octahedra are ubiquitous in the perovskite family, while the appearance 
of breathing distortions -- oxygen displacement patterns that lead to approximately uniform dilation and contraction of the $B$O$_6$ octahedra -- are rarer in compositions with a single, chemically unique $B$-site.
The presence of a CBD relies on electronic instabilities 
of the $B$-site cations, either orbital degeneracies or valence-state fluctuations,  
and often appear concomitant with charge order metal--insulator transitions or 
$B$-site cation ordering.
We enumerate the structural variants obtained from rotational 
and breathing lattice modes and formulate a general 
Landau functional describing their interaction. 
We use this information and combine it with statistical correlation techniques 
to evaluate the role of atomic scale distortions on the critical temperatures 
in representative charge ordering nickelate and bismuthate  perovskites. 
Our results provide  microscopic insights into the underlying structure--property 
interactions across  electronic and magnetic phase boundaries, suggesting 
plausible routes to tailor the behavior of functional oxides by design.

\end{abstract}
\maketitle

\section{Introduction}
Perovskite oxides with chemical formula $AB$O$_3$ and $B$-site transition metal (TM) cations 
exhibit a range of functional electronic transitions that are intimately tied to 
the structure of the fundamental building blocks \cite{Benedek:2012}: 
($i$) the number of unique $B$--O bonds within an octahedron, and 
($ii$) the tilting of corner-connected octahedra.
Adjacent $B$O$_6$ units typically fill space in perovskites through nearly rigid rotations, which produce  deviations of the $B$--O--$B$ bond angles away from the ideal 180$^\circ$ found in the 
cubic aristotype ($Pm\bar{3}m$ symmetry); the rotations are described by two three-dimensional  
irreducible representations (irreps), $M_3^+$ and $R_4^+$, of the high-symmetry structure.\cite{Howard/Stokes:1998}
Combinations of these lattice instabilities -- cooperative bond length distortions and octahedral rotations -- interact  across \emph{structural} phase transitions through elastic stresses and symmetry allowed coupling invariants as described within Landau theory.
Beside changes to crystal symmetry, 
the transition from high temperature (high symmetry) to 
low temperature (low-symmetry) can also produce 
\emph{electronic} metal--insulator (MI) transitions. 
Perovskite oxides with $B$-site cations in 
$t_{2g}^3e_g^1$ ($d^4$), $t_{2g}^6e_g^1$ ($d^7$) and 
$t_{2g}^6e_g^3$ ($d^9$) electronic configurations are particularly susceptible, 
because the low-energy electronic structure  is dictated by 
the octahedral crystal-field split antibonding $e_g$ 
orbitals---the atomic-like $d$-states that are spatially directed at the coordinating oxygen ligands.
MI-transitions which occur simultaneously  with lattice distortions are common in low-dimensional 
materials, e.g.\ Peierls systems.\cite{Canadell:Book} 
In three-dimensional perovskite oxides, however, the most familiar electronic transitions with concomitant changes in the $B$--O bond lengths and octahedral rotations result  from cooperative first-order Jahn-Teller effects: Tetragonal elongations of the $B$O$_6$ octahedra occur to remedy the orbital degeneracies  that \emph{localized} electrons encounter for particular $B$-site cation configurations. And through these distortions, the crystal maintains a uniform TM valence among all $B$-sites across the transition.\cite{Millis/Littlewood/Shraiman:1995}
The Jahn-Teller distortions are described by irreps $\Gamma_3^+$, $M_2^+$ and $R_3^+$, 
and their interaction with octahedral rotations are well-established.\cite{Carpenter/Howard:2009a,*Carpenter/Howard:2009b}
Unusually high-valence states or  $e_g^1$ electrons in 
\emph{delocalized} band states will also produce structural distortions, but in most cases will preserve the uniformity of the $B$--O bonds and the octahedral crystal field of the $B$O$_6$ units in the process.\cite{Goodenough/Rivadulla:2005}
The $B$ cations will readily adopt mixed valence configurations, 
e.g.\ doped perovskites manganites\cite{Salamon/Marcelo:2001} (containing nominally both Mn$^{3+}$ 
and Mn$^{4+}$) and also stoichiometric nickelates\cite{Goodenough:2004_rev} (Ni$^{3+}$) and ferrates\cite{Matsuno/Fujimori_et_al:2002} (Fe$^{4+}$).
The TM  cation will not maintain  an integer valence state ($n$) uniformly on all $B$-sites,  
but rather charge disproportionate (CDP); the simplest case being into two sites as
\[
2\mathrm{B}^{n+}\rightleftharpoons 
\underset{\textrm{site 1}}{\mathrm{B}^{(n-\delta)+}} + 
\underset{\textrm{site 2}}{\mathrm{B}^{(n+\delta)+}}\, , 
\]
where $\delta$ is a fraction of an electron transferred between $B$-sites. 
Electronic correlation and on-site Coulomb repulsion effects will prefer to order the valence
 deviations $\delta$ so that the 
inequivalent lattice sites, 1 and 2, form a periodic arrangement, changing the translational symmetry.
This so-called \emph{charge ordering} (CO) lowers the potential energy of the crystal and gaps the Fermi surface.
\footnote{
The MI-transition results from  electronic localization, which can be intrinsic and electrostatic in nature (electron--electron repulsion), or extrinsic, due to defects or disorder.
}

\begin{figure}
   \centering
   \includegraphics[width=0.9\columnwidth,clip]{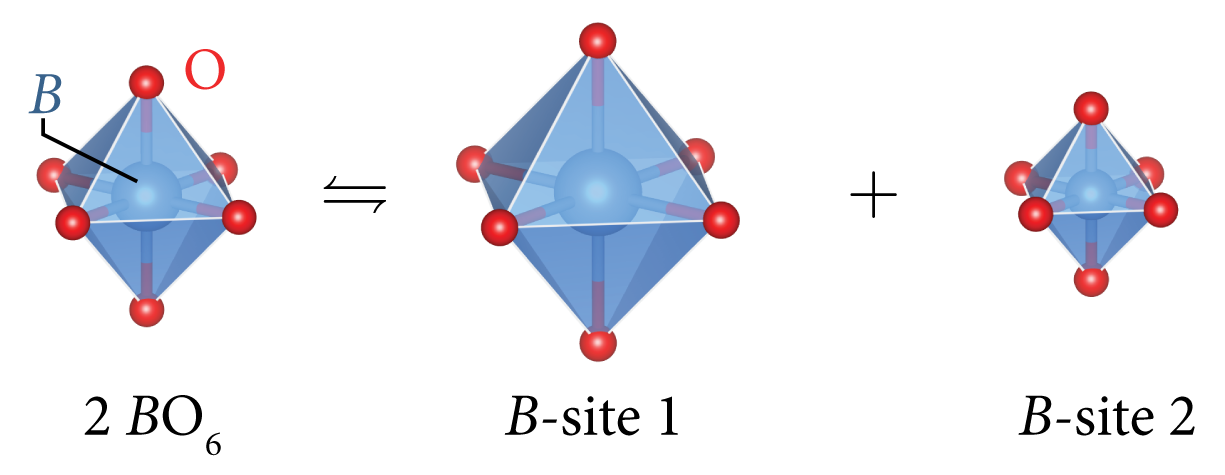}\vspace{-11pt}
   \caption{
   \label{fig:bdp}(Color online.) Illustration of the effect of charge disproportionation (CDP) on the local $B$O$_6$ octahedral site equivalence. Before the CDP transition all octahedra are equivalent (left); afterwards the $B$O$_6$ octahedra disproportionate into non-equivalent sites with the simplest two-site case shown here:  charge transfer between $B$-sites causes one octahedron to dilate (site 1) and the other octahedron to contract (site 2), resembling a ``breathing'' mode of the $B$O$_6$ perovskite building blocks.}
\end{figure}

The electronic charge ordering is 
distinguishable\cite{PhysRevLett.109.216401,*PhysRevLett.109.156402,*PhysRevB.86.195144} 
by an associated structural change in the local $B$O$_6$ building blocks (\autoref{fig:bdp}). It appears as a  $B$--O bond disproportionation or ``breathing distortion,'' which causes the octahedra to either dilate (site 1) or contract (site 2) according  to the charge $\delta$ transferred between sites, 
\footnote{Not all charge-ordering transitions require an concomitant change in structure, see for example the case of LaSr$_2$Fe$_3$O$_{12}$.} 
%
largely because of the change in ionic radii, \emph{i.e.},  the effective radius of $\textrm{B}^{(n-\delta)+}$ is larger than 
$\textrm{B}^{(n+\delta)+}$.

The magnitude of the octahedral rotations are modified by these changes in the $B$-O bond lengths.
The extent to which cooperation between octahedral breathing and rotation instabilities is necessary to stabilizing charge order and MI-transitions, however, is not as well understood \cite{Matsuno/Fujimori_et_al:2002,Saha-Dasgupta/Popovic/Satpathy:2005} 
as the role of Jahn-Teller distortions on electronic transitions \cite{Mizokawa/Khomskii/Sawatzky:1999,Carpenter/Howard:2009a,Carpenter/Howard:2009b}.
To this end, group theoretical methods are particularly powerful to address the interactions among the multiple octahedron-derived instabilities.
They provide a rigorous means to evaluate the symmetry allowed interactions between 
coupled lattice degrees of freedom, and thus,  glean insight into the 
microscopic atomic structural contributions in the electronic CO transition in perovskite oxides.
In this work, we enumerate the space group and order parameter relationships for the octahedral 
breathing distortions, which are associated with irreps $M_1^+$ and $R_1^+$ of the aristotype 
cubic phase and the 15 simple 
octahedral rotation patterns available to bulk $AB$O$_3$ perovskites.
We provide a list of symmetries that perovskite oxides with charge order tendencies 
and octahedral rotations could adopt---not all linear combinations of the instabilities are 
anticipated to be symmetry allowed.\footnote{The list is for $B$-site ordered perovskites where breathing distortions is not due to the spontaneous electron instability, but present because of ordered arrangement of B-site element.}
Using this information, we then illustrate how to quantify relative contributions of octahedral breathing distortions 
and rotations across phase boundaries in prototypical nickelate and bismuthate perovskites. 
This rigorous mapping of the unit-cell level structural distortions into a symmetry-adapted basis enables us 
to disentangle the role that the \emph{atomic structure} plays in directing the macroscopic  
\emph{electronic} metal--insulator transitions. 
Finally, we show that a synergistic combination of group-theoretical analysis with statistical analysis makes it possible to understand the complex interplay pervasive in mixed-metal perovskite oxides.

\section{Cooperative Breathing Distortions}
\begin{figure}
   \centering
   \includegraphics[width=0.98\columnwidth,clip]{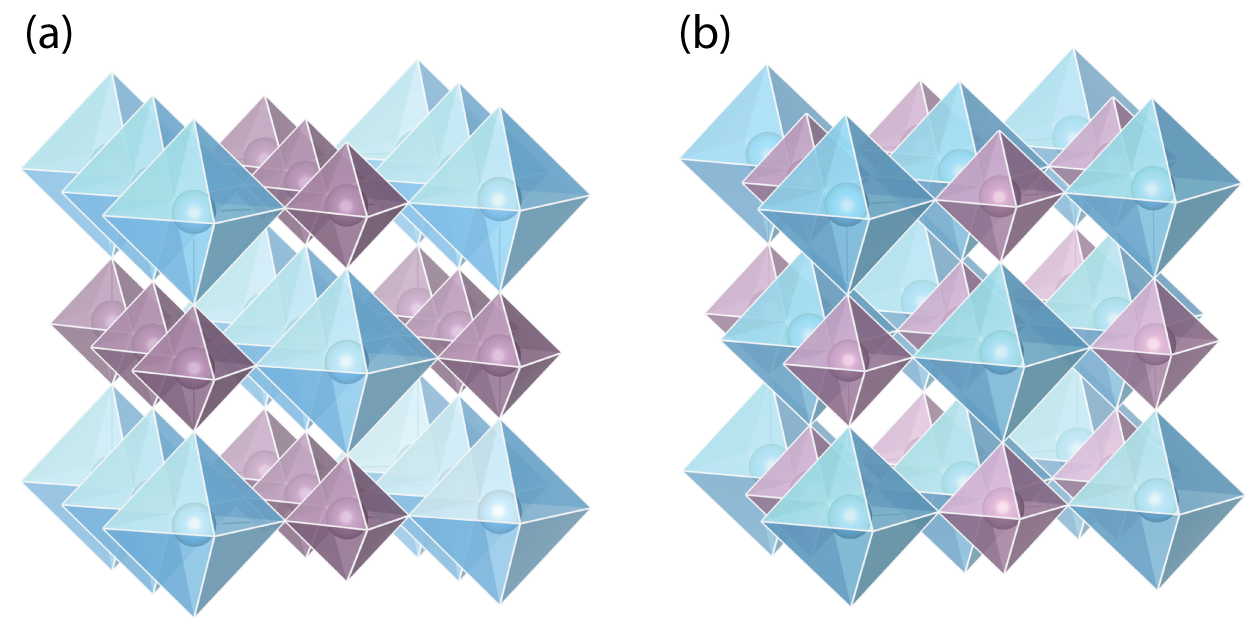}\vspace{-8pt}
   \caption{
   \label{fig:cooperative_breathing}(Color online.) %
Schematic octahedral representations of the two cooperative structural breathing distortions 
that coexists with charge ordering in $AB$O$_3$ perovskites. In (a) the breathing distortion is two-dimensional, irrep $M_{1}^{+}$, and produces a columnar ordering of the two $B$ cations reducing the 
symmetry to $P4/mmm$. In (b) a three-dimensional ordering produces a checkerboard arrangement of the $B$-cations and is described by the $R_1^+$ irrep ($Fm\bar{3}m$ symmetry).}
\end{figure}

The octahedral breathing distortions, which create two unique crystallographic $B$-sites, must 
tile in three-dimensions to maintain the corner-connectivity of the $B$O$_6$ framework---the 
defining feature of the perovskite crystal structure.
Here we consider $B$--O distortions with wavevectors commensurate with the $Pm\bar{3}m$ 
lattice periodicity, \emph{i.e.}, those modes which occur on the edges and corner of the 
simple cubic Brillouin zone.
The two main \emph{cooperative} breathing distortions (CBD) are illustrated in \autoref{fig:cooperative_breathing}.
Note that the irreps, which we describe next, are defined using the $Pm\bar{3}m$ setting of the 
$AB$O$_3$ perovskite with the $B$ cation located at the origin.
The first type of CBD consists of two $B$O$_6$ building blocks which are tiled to form a columnar arrangement of dilated and contracted octahedra [Fig.~\ref{fig:cooperative_breathing}(a)], that splits the $B$--O bond lengths into a doublet and quartet. \footnote{This distortion 
is not a Jahn-Teller elongation of the octahedra, because the equatorial oxygen of a single 
octahedron either all elongate (contract) in the same direction; a Jahn-Teller distortion is
rather seen as a two-in-two-out distortion of the equatorial oxygen atoms.}
This distortion is associated with the active three-dimensional irrep $M_1^+$ with order parameter $(a,0,0)$ and 
manifests as a zone edge $k=(\frac{1}{2},\frac{1}{2},0)$ lattice instability of the cubic phase.
The order parameter (OP) describes a vector in the irrep space and corresponds to specific directions along which the physical distortion may be induced.
For the $M_1^+$ irrep, the OP can have three general components ($a,b,c$), where the values 
$a$, $b$ and $c$ correspond to amplitudes of the two-dimensional breathing distortion along each Cartesian 
direction $x$, $y$ and $z$, respectively.
Here we consider only the case where $b=c=0$, i.e.\ a restricted one-dimensional space.
The possible directions for the OP correspond to $(a,0,0)$, $(0,a,0)$ and $(0,0,a)$ or columnar arrangements 
of the CBD along the $x$-, $y$- and $z$-directions, respectively, so that the 
octahedra distort in the same sense along the given direction.
As a result, the symmetry is reduced to tetragonal, $P4/mmm$ (space group no.\ 123), and the $B$-site Wyckoff position of the cubic aristotype is split as $1a \rightarrow 1a + 1c$, doubling the number of perovskite formula units (f.u.) in the primitive cell (Table~\ref{tab:irreps_structures}). 

The second CBD consists of $B$O$_6$ octahedra which are tiled in a 
three-dimensional checkerboard arrangement   [Fig.~\ref{fig:cooperative_breathing}(b)].
The distortion is described by a one-dimensional irrep $R_4^+$ and occurs as 
a $k=(\frac{1}{2},\frac{1}{2},\frac{1}{2})$ zone-corner lattice instability.
Consequently, in the absence of other distortions, the CBD lifts the $B$-site equivalence 
while maintaining the $O_h$ symmetry of the octahedra through uniform contraction and elongation of the $B$--O bonds about the trigonal axis, splitting the $B$-site Wyckoff position and forming a larger unit cell  (Table~\ref{tab:irreps_structures}).
We note that this type of $B$-site ordering is common in double perovskites with multiple $B$ cations that show large ionic size and/or considerable oxidation state differences, \emph{e.g.}, Ba$_2$MgWO$_6$. \cite{Graham/Woodward:2010} But here, we consider only single (chemical species) $B$ cations with different nominal valences.

\begin{table}
\begin{ruledtabular}
\centering
\caption{\label{tab:irreps_structures} Crystallographic data including the occupied Wyckoff positions (Wyck.\ Site) 
for the cooperative breathing distortions (CBD) available to $AB$O$_3$ perovskites (\autoref{fig:cooperative_breathing}) in the absence of $B$O$_6$ rotations. 
Atom positions are given relative to the ideal cubic symmetry such that the relevant 
CBD imposed on the oxygen positions is indicated by $\Delta$. The value of $\Delta$ 
controls the amplitude of the $B$--O contraction (elongation) 
and typically scales with the amount of inter-site charge  transfer ($\delta$). The change in cell size 
is given relative to the pseudo-cubic (pc) lattice constant $a_\textrm{pc}$.}
\begin{tabular}{lcccc}%
\multicolumn{5}{l}{$M_1^+$ $(a,0,0)$ \hfill  $a=b\sim \sqrt{2}a_\textrm{pc},  c\sim a_\textrm{pc}$}  \\
\multicolumn{5}{l}{$P4/mmm$ (no.\  123) \hfill $\alpha=\beta=\gamma=90^\circ$}\\[0.4ex]
Atom	&	Wyck.\ Site	& $x$	& $y$	& $z$ 	\\
\hline
$A$    	&$2e$&  0 		& $\frac{1}{2}$	&	$\frac{1}{2}$ \\
$B$(1)    &$1a$& 0		& 0				& 0				\\
$B$(2)    &$1c$& $\frac{1}{2}$	&	$\frac{1}{2}$	& 0	\\
O(1)       &$4j$& $\frac{1}{4} + \Delta$ & $\frac{1}{4} + \Delta$ & 0 \\
O(2)       &$1b$& 0		& 0				&	$\frac{1}{2}$\\
O(3)       &$1d$& $\frac{1}{2}$	&	$\frac{1}{2}$	& $\frac{1}{2}$ \\[0.2em]
\hline\\[-0.8em]
\multicolumn{5}{l}{$R_1^+ (a)$ \hfill  $a=b=c\sim 2a_\textrm{pc}$}  \\
\multicolumn{5}{l}{$Fm\bar{3}m$ (no.\ 225) \hfill $\alpha=\beta=\gamma=90^\circ$}\\ [0.4ex]
Atom	&	Wyck.\ Site	& $x$	& $y$	& $z$ 	\\
\hline
$A$		& $8c$ 	&$\frac{1}{4}$	&	$\frac{1}{4}$	& $\frac{1}{4}$\\
$B$(1) 	& $4a$ 	& 0		& 0				& 0				\\
$B$(1) 	& $4b$ 	&$\frac{1}{2}$	&	$\frac{1}{2}$	& $\frac{1}{2}$\\
O  		& $24e$ 	&$\frac{1}{4} + \Delta$ &0  & 0 \\
\end{tabular}
\end{ruledtabular}
\end{table}

\section{Octahedral Rotation and CBD Space Groups}

\subsection{Methodology}
To enumerate the allowed combinations of CBD and octahedral rotations in 
perovskite oxides we use the group theoretical program \textsc{isotropy}.\cite{Campbell/Stokes/Tanner/Hatch:2006}
We follow the approach of Stokes \emph{et al.}\cite{Stokes/Howard:2005} 
and consider the changes in lattice symmetry due to the 
superposition of each CBD pattern (\autoref{fig:cooperative_breathing}) with the 15 octahedral 
rotation systems derived from the $M_3^+$ and $R_4^+$ three-dimensional irreps describing rotations.\cite{Howard/Stokes:1998}
The advantage of this approach is that group--subgroup relationships can be established between 
structural variants, enabling the understanding of the structural and electronic CO transitions 
within a first- or second-order theory. It should be noted that the analysis of \emph{charge} order at 
a particular site in the crystal is effectively the same as analyzing the effects of \emph{cation} order due to a 
order--disorder transition, because both valence and chemical species split a Wyckoff position identically. In the 
latter case, the irrep describes a ``composition'' mode and reflects the site occupancy.
In this work, we enumerate those mode combinations 
with well-defined cooperative breathing distortions that 
are most likely to be observed experimentally.
Although the rock salt $B$ cation order has been studied previously in 
$AB$O$_{3}$ perovskites, the layered ordering of $B$ cations, \emph{i.e.} irrep $M_{1}^{+}$, 
was examined only for the compositionally complex quadruple perovskite 
$A_{4}BB_{3}^{\prime}$O$_{12}$ oxides.\cite{Howard/Stokes:2004}
Here, we report the results of the $M_{1}^{+}$ irrep on simple $AB$O$_{3}$ perovskites in the context 
of the CBD behavior, where cell-doubling occurs.
Specifically for the $M_1^+$ irrep, we retain only those structures with an OP 
$(a,0,0)$ such that the second and third vector components are zero, unless an octahedral 
rotation permits it to be non-zero by symmetry.
Furthermore, we follow the convention introduced in Ref.~\onlinecite{Howard/Stokes:1998,Stokes/Kisi_et_al:2002} 
and only keep structures with coherent rotations---those with a fixed  ``sense'' about each axis.
We remove structure variants from our analysis which would allow for a modulation in the 
amplitude and sense of the rotations about a single axis.

\subsection{Space Groups}
We enumerate the space groups (Table \ref{tab:cdp}) and associated octahedral 
tilt pattern, irreps, lattice vectors, and origin allowed by group theory. 
We follow Glazer's established notation \cite{Glazer:1972} to denote the 
magnitude and phase of octahedral tilt patterns in perovskites: 
The description of octahedral rotations are encoded using the syntax 
$a^{\#}b^{\#}c^{\#}$, where letters $a$, $b$, and $c$ indicate rotations of 
$B$O$_6$ units of equal or unequal magnitude about Cartesian $x$-, $y$-, and $z$-axes.
Note that in the case of equal magnitude rotations about different axes, the equivalent letter 
is duplicated, \emph{e.g.}, $a^{\#}a^{\#}a^{\#}$.
The superscript $\#$ can take three values: 0, +, or $-$, for no rotations, 
in-phase rotations (neighboring octahedra along a Cartesian axis rotate in the same direction), 
or out-of-phase rotations (adjacent octahedra rotate in the opposite direction), respectively. 

\begingroup
\squeezetable
\begin{table*}[]
\begin{ruledtabular}
\centering
\caption{\label{tab:cdp}Possible crystallographic space groups, octahedral rotation patterns and unit cell 
relationships for $AB$O$_3$ perovskites exhibiting rotations of octahedra given by 
irreps $M_3^{+}$ and $R_4^{+}$ with order parameter directions given in 
parentheses ($\eta_i,\eta_j,\eta_k$) with the CBD $M_1^+$ or $R_1^+$.
The lattice vectors and origin shifts are given with respect to the high 
symmetry 5-atom $Pm\bar{3}m$ structure ($B$ cation at the origin).}
\resizebox{1.90\columnwidth}{!}{
\begin{tabular}{llcccll}
			&			&\multicolumn{3}{c}{Order parameter direction}	&	& \\
\cline{3-5}
space group  	&  tilt pattern 	&  & $M_3^+$ & $R_4^+$ 	& lattice vectors 		& origin \\
\hline
221 $Pm\bar{3}m$& $a^0a^0a^0$  & &(0,0,0) &(0,0,0)	    &$(1,0,0),(0,1,0),(0,0,1)$  	    & $(0,0,0)$     \\[0.8ex]
127 $P4/mbm$& $a^0a^0c^+$      & &(a,0,0) &(0,0,0)	    &$(1,1,0),(\bar{1},1,0),(0,0,1)$	    & $(0,0,0)$     \\
139 $I4/mmm$& $a^0b^+b^+$      & &(a,0,a) &(0,0,0)	    &$(0,2,0),(0,0,2),(2,0,0)$  	    & $(\frac{1}{2},\frac{1}{2},\frac{3}{2})$	    \\
204 $Im\bar{3}$& $a^+a^+a^+$   & &(a,a,a) &(0,0,0)	    &$(2,0,0),(0,2,0),(0,0,2)$  	    & $(\frac{1}{2},\frac{1}{2},\frac{1}{2})$	    \\
71  $Immm$& $a^+b^+c^+$        & &(a,b,c) &(0,0,0)	    &$(2,0,0),(0,2,0),(0,0,2)$  	    & $(\frac{1}{2},\frac{1}{2},\frac{1}{2})$	    \\[0.8ex]
140 $I4/mcm$& $a^0a^0c^-$      & &(0,0,0) &(a,0,0)	    &$(1,1,0),(\bar{1},1,0),(0,0,2)$	    & $(0,0,0)$     \\
74 $Imma$& $a^0b^-b^-$ 	       & &(0,0,0) &(a,0,a)	    &$(0,1,1),(2,0,0),(0,1,\bar{1})$	    & $(0,0,0)$     \\
167 $R\bar{3}c$	& $a^-a^-a^-$  & &(0,0,0) &(a,a,a)	    &$(\bar{1},1,0),(0,\bar{1},1),(2,2,2)$ & $(0,0,0)$      \\
12 $C2/m$& $a^0b^-c^-$ 	       & &(0,0,0) &(a,0,b)	    &$(0,\bar{2},0),(2,0,0),(0,1,1)$	    & $(\frac{1}{2},\frac{1}{2},0)$ \\
15 $C2/c$& $a^-b^-b^-$ 	       & &(0,0,0) &(a,b,a)	    &$(2,\bar{1},\bar{1}),(0,1,\bar{1}),(0,1,1)$& $(\frac{1}{2},\frac{1}{2},0)$     \\
2 $P\bar{1}$& $a^-b^-c^-$      & &(0,0,0) &(a,b,c)	    &$(0,1,1),(1,0,1),(1,1,0)$  	    & $(0,0,0)$     \\[0.8ex]
63 $Cmcm$& $a^0b^+c^-$ 	       & &(0,0,a) &(b,0,0)	    &$(2,0,0),(0,0,\bar{2}),(0,2,0)$	    & $(\frac{1}{2},0,\frac{1}{2})$ \\
62 $Pnma$& $a^+b^-b^-$ 	       & &(0,a,0) &(b,0,b)	    &$(0,1,1),(2,0,0),(0,1,\bar{1})$	    & $(0,0,0)$     \\
11 $P2_1/m$& $a^+b^-c^-$       & &(0,a,0) &(b,0,c)	    &$(0,\bar{1},1),(2,0,0),(0,1,1)$	    & $(0,0,0)$     \\
137 $P4_2/nmc$& $a^+a^+c^-$    & &(0,a,a) &(b,0,0)	    &$(2,0,0),(0,2,0),(0,0,2)$  	    & $(0,0,\bar{1})$	    \\[1.2ex]
\hline
&  &  $M_1^+$ & & & & \\ 
\cline{3-3}\\[-1em]
\textbf{127 \emph{P}4/\emph{mbm}}& $a^0a^0c^0$\footnote{In the absence of octahedral rotations, the axes are unique due to the CBD.}      & (a,0,0)\footnote{The CBD makes the $z$-direction unique.} &(0,0,0)&(0,0,0)	    &$(1,1,0),(\bar{1},1,0),(0,0,1)$	    & $(0,0,0)$     \\
83 $P4/m$	&	$a^0a^0c^+$	&	(a,0,0) &(b,0,0)&(0,0,0)	    &$(1,1,0),(\bar{1},1,0),(0,0,1)$	    & $(0,0,0)$     \\
74 $Imma$	&	$a^+a^0c^0$	&	(a,0,0) &(0,0,b)&(0,0,0)	    &$(0,2,0),(0,0,2),(2,0,0)$	    & $(0,0,0)$     \\
69 $Fmmm$	&	$a^+a^+c^0$	&	(a,0,0) &(0,b,b)&(0,0,0)	    &$(0,\bar{2},0),(2,0,2),(\bar{1},1,1)$	    & $(\frac{1}{2},-\frac{1}{2},-\frac{1}{2})$     \\
12 $C2/m$	&	$a^+b^+c^+$	&	(a,0,0) &(b,c,d)&(0,0,0)	    &$(\bar{2},2,0),(0,0,\bar{2}),(0,2,0)$	    & $(\frac{1}{2},-\frac{1}{2},\frac{1}{2})$     \\[0.8ex]
124 $P4/mcc$	&	$a^0a^0c^-$	&	(a,0,0) &(0,0,0)&(b,0,0)	    &$(1,1,0),(\bar{1},1,0),(0,0,1)$	    & $(0,0,0)$     \\
63 $Cmcm$	&	$a^-a^0c^0$	&	(a,0,0) &(0,0,0)&(0,0,b)	    &$(0,2,0),(0,0,2),(2,0,0)$	    & $(0,0,0)$     \\
51 $Pmma$	&	$a^-a^-c^0$	&	(a,0,0) &(0,0,0)&(0,b,b)	    &$(0,0,2),(1,1,0),(\bar{1},1,0)$	    & $(0,0,0)$     \\
13 $P2/c$	&	$a^-a^-c^-$	&	(a,0,0) &(0,0,0)&(c,b,b)	    &$(1,1,0),(\bar{1},1,0),(0,0,2)$	    & $(0,0,0)$     \\
15 $C2/c$	&	$a^0b^-c^-$	&	(a,0,0) &(0,0,0)&(b,c,0)	    &$(2,0,0),(0,2,0),(0,0,2)$	    & $(0,0,0)$     \\
11 $P2_1/m$	&	$a^-b^-c^0$	&	(a,0,0) &($b^\prime$,0,0)\footnote{The $b^\prime$ denotes that this is a secondary distortion that is symmetry allowed. In this case in-phase rotations about the $z$-direction. The Glazer tilt is defined for the case $b^\prime=0$.}&(0,c,d)	    &$(\bar{1},1,0),(0,2,0),(1,1,0)$	    & $(0,0,0)$     \\
2 $P\bar{1}$	&	$a^-b^-c^-$	&	(a,0,0) &($b^\prime$,0,0)$^\textrm{c}$&(c,d,e)	    &$(1,1,0),(\bar{1},1,0),(0,0,2)$	    & $(0,0,0)$     \\[0.8ex]
52 $Pnna$	&	$a^0b^+c^-$	&	(a,0,0) &(0,0,b)&(c,0,0)	    &$(0,0,2),(0,2,0),(\bar{2},0,0)$	    & $(0,0,0)$     \\
62 $Pnma$	&	$a^+b^-c^0$	&	(a,0,0) &(0,0,b)&(0,c,0)	    &$(0,2,0),(0,0,2),(2,0,0)$	    & $(0,0,0)$     \\
11 $P2_1/m$	&	$a^-b^-c^+$		&	(a,0,0) &(b,0,0)&(0,c,d)	     &$(\bar{1},1,0),(0,2,0),(1,1,0)$	    & $(0,0,0)$     \\
68 $Ccca$	&	$a^+a^+c^-$	&	(a,0,0) &(0,b,b)&(c,0,0)	    &$(2,2,0),(\bar{2},2,0),(0,0,2)$	    & $(2,0,0)$     \\[1.2ex]
\hline
&  &  $R_1^+$ & & & & \\
\cline{3-3}\\[-1em]
\textbf{225 \emph{Fm}$\bar{3}$\emph{m}}	&	$a^0a^0a^0$\footnote{The $R_1^+$ mode acts along the body diagonal and does not create a unique axis in the crystal as in the case of the the three-dimensional $M_1^+$ irrep. Only the octahedral rotations lifts the axes equivalence.}		&	(a) &(0,0,0)&(0,0,0)	    &$(2,0,0),(0,2,0),(0,0,2)$	    & $(0,0,0)$     \\
128 $P4/mnc$	&	$a^0a^0c^+$&	(a) &(b,0,0)&(0,0,0)	    &$(1,1,0),(\bar{1},1,0),(0,0,2)$	    & $(0,0,0)$     \\
201 $Pn\bar{3}$	&	$a^+a^+a^+$	&	(a) &(b,b,b)&(0,0,0)	     &$(0,0,\bar{2}),(0,\bar{2},0),(\bar{2},0,0)$	 	    & $(1,1,1)$     \\
134 $P4_2/nnm$	&	$a^0b^+b^+$	&	(a) &(b,b,0)&(0,0,0)	     &$(0,0,2),(2,0,0),(0,2,0)$	 	    & $(0,\bar{1},0)$     \\
48 $Pnnn$	&	$a^+b^+c^+$	&	(a) &(b,c,d)&(0,0,0)	     &$(\bar{2},0,0),(0,0,2),(0,2,0)$	 	    & $(0,0,0)$     \\[0.8ex]
148 $R\bar{3}$	&	$a^-a^-a^-$	&	(a) &(0,0,0)&(b,b,b)	    &$(\bar{1},1,0),(0,\bar{1},1),(2,2,2)$	    & $(0,0,0)$     \\
87 $I4/m$	&	$a^0a^0c^-$	&	(a) &(0,0,0)&(b,0,0)	    &$(\bar{1},1,0),(\bar{1},\bar{1},1),(0,0,2)$	    & $(0,0,0)$     \\
12 $C2/m$	&	$a^0b^-b^-$	&	(a) &(0,0,0)&(b,b,0)	    &$(\bar{1},2,1),(1,0,1),(0,2,0)$	    & $(0,0,0)$     \\
2 $P\bar{1}$	&	$a^-b^-c^-$	&	(a) &(0,0,0)&(b,c,d)	    &$(1,0,1),(1,1,0),(\bar{1},1,0)$	    & $(0,0,0)$     \\[0.8ex]
15 $C2/c$	&	$a^0b^+c^-$	&	(a) &(b,0,0)&(0,0,c)	    &$(2,0,0),(0,2,0),(0,0,2)$	    &  $(\frac{1}{2},\frac{1}{2},0)$     \\
86 $P4_2/n$	&	$a^+a^+c^-$	&	(a) &(b,b,0)&(0,0,c)	   &$(0,0,2),(2,0,0),(0,2,0)$		    & $(\bar{1},\bar{1},0)$     \\
14 $P2_1/c$	&	$a^+b^-b^-$		&	(a) &(b,0,0)&(0,c,c)	    &$(1,\bar{1},0),(1,1,0),(1,\bar{1},2)$	    & $(0,0,0)$     \\
\end{tabular}
}
\end{ruledtabular}\vspace*{0.2em}
\end{table*}
\endgroup

\autoref{tab:cdp} provides the possible space group symmetries compatible with 
CBD and octahedral rotations. 
The structural data is divided into three main blocks: The first section contains the 
space group symmetries in the absence of CBD; the second section enumerates the symmetries that result from the planar CBD with octahedral rotations; and the third section 
includes those obtained from the three-dimensional CBD combined with rotations.
Structures appearing in bold in \autoref{tab:cdp} 
correspond to CBD without any octahedral rotations and are given first 
at the top of block two and three.
Note that for perovskites with rotations and the $M_1^+$ CBD, there may be 
more than one structure possible for a given rotation, because the relative orientation of the tilt pattern with respect to the columnar arrangement of dilated and contracted octahedra alters the crystal symmetry differently.
\autoref{tab:cdp} directly reveals the effect of the superposition of 
octahedral rotation patterns and the CBD patterns on the crystal symmetry lowering. 
Consider the middle block ($M_{1}^{+}$$\oplus$$M_{3}^{+}$$\oplus$$R_{4}^{+}$). The seemingly similar tilt patterns $a^{0}a^{0}c^{+}$ and $a^{+}a^{0}c^{0}$,  depending on the crystallographic axes they act upon, 
yield two different space groups, $P2/m$ and $Imma$, respectively. 

Furthermore, there are a number of space groups appearing in each of the blocks of 
\autoref{tab:cdp}: $C2/m$ appears both in the rotation only block (\emph{sans} any  
CBD) and in the third block with the $R_1^+$ CBD. 
Formally the crystallographic symmetries are identical in each case; however, physically 
the rotation patterns adopted by the crystals are different. 
For instance, when only rotations associated with irrep $R_4^+$ and order parameter $(a,0,b)$ are present in the 
crystal structure, space group $C2/m$ allows 
the out-of-phase rotations about the $y$- and $z$-axes to be of different magnitude (tilting pattern is $a^0b^-c^-$).
Now, consider the direct sum $R_1^+$$\oplus$$R_4^+$ for a vector space corresponding to the order parameter $(a,b,b,0)$ that includes CBD. Even though the resulting space group is $C2/m$, the corresponding octahedral tilt pattern changes to $a^0b^-b^-$, where the out-of-phase rotations about the $y$- and $z$-axes are of the same magnitude. 
On the other hand, the direct sum $M_1^+$$\oplus$$R_4^+$ for a vector space corresponding to the order parameter $(a,0,0,b,c,0)$ is found to preserve the $a^0b^-c^-$ tilting pattern, but the overall symmetry reduces to $C2/c$.
Also note that, in the case of rotations without CBD, the tilt pattern $a^0b^-b^-$ corresponds to the higher symmetry $Imma$ space group.
Such restrictions imposed on the rotation axes equivalence 
are described 
next, and for the reason previously described explains why ``none'' appears as an 
entry in \autoref{tab:tilts_co}.

\subsection{Compatible CBD and Rotation Symmetries}

\begin{figure}[b]
\centering
   \includegraphics[width=0.88\columnwidth,clip]{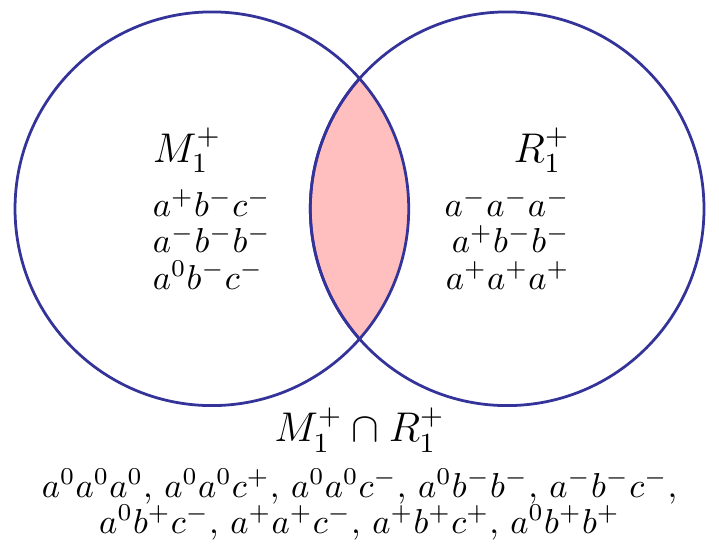}\vspace{-8pt}
\caption{\label{fig:venn_diagram}(Color online.) %
Representation of the rotation patterns that are compatible with each 
(or both, given by the intersection of the) CBD without 
requiring a further symmetry reduction.}
\end{figure}

\begingroup
\squeezetable
\begin{table}[t]
\begin{ruledtabular}
\centering
\caption{\label{tab:tilts_co}The change in space group which occurs when the 
different octahedral CBD patterns are superposed with the 15 simple octahedral tilt systems.} 
\begin{tabular}{llll}
Tilt system		&	&	& \\
(Glazer notation)	&	No Breathing	&	$M_1^+$	& $R_1^+$ \\[0.2em]
\hline
$a^0a^0a^0$				&	$Pm\bar{3}m$ &	$P4/mmm$	&	$Fm\bar{3}m$ \\
\hline
$a^0a^0c^+$				&	$P4/mbm$ &	$P4/m$	&	$P4/mnc$ \\
	&	&														$Imma$ 	&						\\
\hline
$a^0b^+b^+$				&	$I4/mmm$ &	$Fmmm$	&	$P4_2/nnm$ \\	
$a^+a^+a^+$				&	$Im\bar{3}$ &	none	&	$Pn\bar{3}$ \\
$a^+b^+c^+$				&	$Immm$ &	$C2/m$	&	$Pnnn$ \\
\hline
$a^0a^0c^-$				&	$I4/mcm$ &	$P4/mcc$	&	$I4/m$ \\
	&	&														$Cmcm$ 	&						\\
\hline
$a^0b^-b^-$				&$Imma$		&	 $Pmma$		&$ C2/m$	\\
$a^-a^-a^-$				&$R\bar{3}c$	&	none	&	$R\bar{3}$ \\
$a^0b^-c^-$				&$C2/m$		&	$C2/c$ &	none \\
					& &			$P2_1/m$	& \\
\hline
$a^-b^-b^-$				&$C2/c$		&		$P2/c$	& none	\\
$a^-b^-c^-$				&$P\bar{1}$	&	no change	&	no change \\
\hline
$a^0b^+c^-$				&$Cmcm$	&	$Pnna$	& $C2/c$ \\
									& & 		$Pnma$	&	\\
\hline
$a^+b^-b^-$	&			$Pnma$			&	none	&	$P2_1/c$\\
$a^+b^-c^-$	&			$P2_1/m$		&$P2_1/m$		& none \\
$a^+a^+c^-$	&			$P4_2/nmc$	&$Ccca$		& $P4_2/n$\\
\end{tabular}
\end{ruledtabular}
\end{table}
\endgroup
In Table \ref{tab:tilts_co}, we aggregate the results of the change in space group symmetry due to the superposition of tilt patterns of $M_{1}^{+}$ and $R_{1}^{+}$ irreps with each of the 15 simple octahedral tilt systems. 
This information is schematically  shown in \autoref{fig:venn_diagram}. 
Only one symmetry exists, $P\bar{1}$ , where the breathing distortion is geometrically 
compatible with the symmetry of the perovskite structure with octahedral rotations alone 
($a^-b^-c^-$), meaning ``no change'' in unit cell or translational symmetry is required 
to accommodate the multiple distortions.
\autoref{tab:tilts_co} also reveals that the $P2_1/m$ symmetry is compatible with two 
different octahedral rotation patterns and the $M_1^+$ CBD: $a^0b^-c^-$ and 
$a^+b^-c^-$.
While $P\bar{1}$ is compatible with both $M_{1}^{+}$ and $R_{1}^{+}$ irreps, $P2_{1}/m$ only exists for the $M_{1}^{+}$  irrep.

For entries containing ``none,'' the direct sums, $R_{1}^{+}$$\oplus$$M_{3}^{+}$$\oplus$$R_{4}^{+}$ and 
$M_{1}^{+}$$\oplus$$M_{3}^{+}$$\oplus$$R_{4}^{+}$, do not yield an isotropy subgroup. 
This is made clear by examining the $a^-a^-a^-$  rotation pattern, which corresponds 
to an equal amplitude of out-of-phase rotations about each axis, or equivalently 
a single out-of-phase rotation about the three-fold axis.
The trigonal symmetry is incompatible with a distortion that would 
require a loss of the three-fold axis and therefore need at a minimum a 
symmetry reduction to a lattice with tetragonal geometry.
Thus, `none' appears in the $M_1^+$ column corresponding to the 
row with the $a^-a^-a^-$ rotation pattern.
Such incompatibility with the  $M_1^+$ CBD is alleviated if the rotation pattern 
about any two crystallographic axes are of different magnitudes, \emph{e.g.},  $a^-b^-b^-$, yielding space 
group $P2/c$ (\autoref{tab:tilts_co}).

\section{Structural Transitions}

Although the microscopic origin for charge disproportionation 
results from an electronic instability related to the electronic 
configuration of a particular metal center, the cooperative ordering 
of the CDP leads to a macroscopic bond-disproportionation. 
In the displacive limit, the two unique and nearly uniform 
octahedra result, dilating and contracting in proportion to the 
magnitude of charge transfer.
Across the electronic phase transitions, the breathing distortions 
can couple directly to octahedral rotations. 
(We do not consider here indirect coupling 
through a common strain component.)

\begin{figure}[b]
\centering
   \includegraphics[width=0.99\columnwidth,clip]{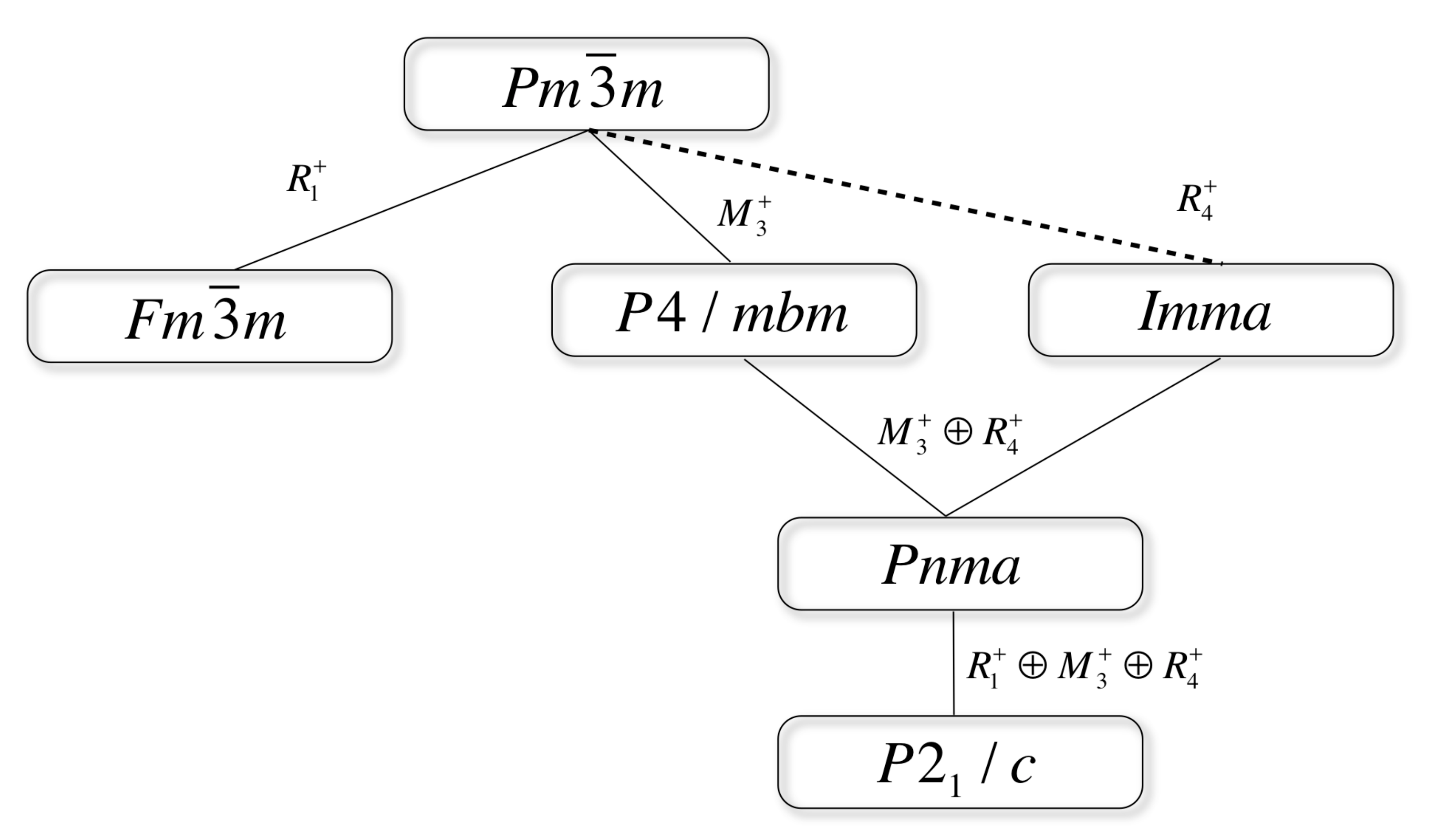}\vspace{-7pt}
    \caption{Group---sub-group relationship between $Pm\bar{3}m$ and $P2_{1}/c$. Irreps within $Pm\bar{3}m$ responsible for  transitions  are shown. In addition, we note that $M_{3}^{+}$$\oplus$$R_{4}^{+}$$\oplus$$R_{3}^{+}$ and $M_{3}^{+}$$\oplus$$R_{4}^{+}$$\oplus$$M_{5}^{+}$ would also transform $Pm\bar{3}m$ $\rightarrow$ $P2_{1}/c$. Dotted line indicates that the phase transition is not allowed to be continuous within the confines of Landau theory.}
    \label{fig:gp_subgp}
\end{figure}

Here we describe the effect of superposition of the common 
orthorhombic rotation pattern ($a^+b^-b^-$) obtained from  
irreps $M_{3}^{+}$ and $R_{4}^{+}$, which gives 
the six-dimensional order parameter with two unique directions as
\[
(\eta_+,0,0,0,\eta_-,\eta_-)\,.
\]
The superposition of the three-dimensional CBD $R_1^+$ and the rotation pattern 
gives a seven-dimensional order parameter space, \emph{i.e.},
($\beta,\eta_+,0,0,0,\eta_-,\eta_-$), which we contract to 
an effective four-dimensional space to obtain the 
$P2_{1}/c$ space group (cf.\ \autoref{tab:cdp}).
Without assuming which transition occurs first---whether the 
rotations precede or follow the CBD---we 
construct a Landau free energy expansion about the cubic phase 
($\beta=\eta_+=\eta_-=0$) as 
\[
\mathcal{F}(\beta,\eta_i) 
= 
\phi^2 + \phi^4 + C_1\beta^2\eta_+^2 + C_2\beta^2\eta_-^2 + D_3\eta_+^2\eta_-^2\,, 
\]
where $ \phi^2$ and $\phi^4$ describe the homogeneous quadratic and 
quartic terms for each order parameter, and  $C_{i}$ are the coefficients coupling the 
CBD to the in- or out-of-phase rotations, and $D_3$ describes the biquadratic 
coupling of the different $B$O$_6$ rotation ``senses'' in the $a^+b^-b^-$ tilt pattern.
The group--subgroup relationships are depicted  
in \autoref{fig:gp_subgp}.
The rotations and the CBD couple biquadratically, which 
indicates that the interactions across the transitions could be either cooperative 
or antagonistic. 
It is possible that the particular rotations could suppress CO by eliminating the structural 
 octahedral breathing distortion altogether through the interaction terms containing the  $C_i$ coefficients.
However, in most cases, the rotation amplitudes are weakly modified across the 
electronic CO transition, suggesting that the strength of the coupling $C_i$ is in 
general small. 
We explore this in the \autoref{sec:applications} through a statistical approach. 
We also note that while the order of the coupling is important, the difference in 
temperature scales at which the rotation and charge ordering occurs is also important in 
determining how the structural order parameters influence each other. 
The strongest interaction occurs when the temperatures are similar, 
while if they are far apart, the two structural transitions will weakly couple.

\section{Structure--Functionality Relationships\label{sec:applications}}

In this section, we apply our group theory results to quantitatively explore the 
relationship between structure and physical properties of experimentally known 
rare-earth nickelate and bismuthate perovskites. 
There is  significant interest in developing strategies 
-- both experimentally and theoretically -- to rationally control octahedral distortions 
through the interplay of chemical pressure, epitaxial strain engineering,  
and ultrathin superlattice heterostructure formation\cite{Torrance/Lacorre/Nazzal/etal:1992, May:2010, Rondinelli/May/Freeland:2012, Chakhalian/Rondinelli/Liu/etal:2011, Blanca/Pancheva:2011, Boris/Matiks/etal:2011} for property control.
Although several octahedral distortion metrics, \emph{e.g.}, 
the crystallographic tolerance factor or bending of the $B$--O--$B$ bond angle,\cite{Zhou/Goodenough/Dabrowski:2005, Catalan:2008}
have played an important role in the understanding of the electronic and magnetic 
properties of perovskite oxides,\cite{Goodenough:itinerant} 
they have had limited success in materials design of non-thermodynamic phases in thin film 
geometries.
Knowledge of  quantitative structure--property octahedral distortion 
relationships are required to accelerate materials discoveries. 

Distortion-mode decomposition analysis is an alternative approach\cite{Campbell/Stokes/Tanner/Hatch:2006, Carpenter/Howard:2009a,Carpenter/Howard:2009b, Perez-Mato/Orobengoa/Aroyo:2010} 
(widely practiced and followed in the crystallography literature) to study displacive\cite{Dove:1997}
phase transitions in perovskites.
It involves describing a distorted (low-symmetry) structure as 
arising from a (high-symmetry) parent structure with one or more 
static symmetry-breaking structural distortions.\cite{Perez-Mato/Orobengoa/Aroyo:2010}
In the undistorted parent structure, each symmetry breaking distortion-mode has zero amplitude. 
The low-symmetry phase, however, will have finite amplitudes for each irrep 
compatible with the symmetry breaking.
Said another way, the low-symmetry phase is rigorously described through a series 
expansion of static symmetry breaking structural modes that ``freeze'' into the parent 
structure.
Critically, the  weights or amplitudes assigned to each irrep are obtained 
according to the contribution that each irrep is present \emph{and} 
the requirement that linearity is maintained. 
What is of particular utility in formulating quantitative relationships connecting 
octahedral distortions, which are now described mathematically, to macroscopic 
properties for materials design is that each irrep carries a physical representation of the 
displacive distortions---the unique atomic coordinates describing various symmetry-adapted structural modes.
The relative importance of these modes on properties may then be mapped by means of 
\emph{ab initio} computational methods. \cite{Dieguez/Gonzalez/etal:2011}
Accessibility to computational methods make the distortion-mode analysis 
powerful, because it is possible to independently study various distortions
and directly assess their role in structural and electronic phase transition mechanisms.
Furthermore, the distortion-mode analysis relies solely on crystal structural data, 
which enables both bulk and thin film stabilized structures with identical compositions 
to be evaluated on equal-footing. 
Such direct comparison is not possible through aggregate parameters such as the 
tolerance factor, \emph{i.e.} when the composition is fixed, or other metrics 
widely followed in the literature.

In the remainder of this  section, we use the distortion-modes to form 
the basis for the quantitative description of octahedral distortions and CBD on 
material properties. 
Using bulk $R$NiO$_{3}$, where $R$ is a rare-earth element,  
and Ba$_{1-x}$K$_{x}$BiO$_{3}$ perovskite compounds 
as prototypical charge-ordering materials, we decompose available low-symmetry structural 
data into symmetry-adapted structural distortion modes. 
We then evaluate and correlate the amplitudes of the distortion-modes to macroscopic 
materials behavior to uncover trends linking octahedral distortions to 
the structural and physical properties. %
We use the group-theory program \textsc{isodistort} \cite{Campbell/Stokes/Tanner/Hatch:2006} for distortion-mode decomposition analysis and  \texttt{R}  \cite{R:2012, R/Lattice:2008} for the statistical analysis. 
For readers interested in reproducing our work, we have deposited the raw data and the \texttt{R}-script in the supplementary materials section available at \href{http://link.aps.org/supplemental/XYZ}{http://link.aps.org/supplemental/XYZ}.
\subsection{\emph{\textbf{R}}NiO$_{3}$ Nickelates}
Rare-earth perovskite nickelates, $R$NiO$_{3}$, where 
$R=$Y, Ho, Er, Dy, Lu, Pr, or Nd, exhibit  non-trivial changes in structure and 
physical properties, including sharp first-order temperature-driven MI-transitions, 
unusual antiferromagnetic order in the ground state, and site- or bond-centered 
charge disproportionation.\cite{Medarde:1997, Alonso:1999, Zhou/Goodenough/Dabrowski:2005} 
At the MIT temperature ($T_{MI}$), the crystal symmetry lowers from orthorhombic $Pnma$ to monoclinic $P2_{1}/c$ symmetry \footnote{The first origin choice is only used for notation purposes to avoid confusion. Our actual mode-decomposition was performed using $P2_{1}/n$ space group as reported in the literature.} (\autoref{fig:gp_subgp}), where the Ni cation no longer maintains a unique uniform valence on all sites, and  disproportionates as
\[
2\mathrm{Ni}^{3+}\rightleftharpoons 
\underset{\textrm{site 1}}{\mathrm{Ni}^{(3-\delta)+}} + 
\underset{\textrm{site 2}}{\mathrm{Ni}^{(3+\delta)+}}\, .
\]
Moreover, the insulating ground-state displays a complex antiferromagnetic order (Type-$E^\prime$) below a N\'{e}el temperature ($T_{N}$). Several previous studies \cite{Medarde/Lacorre/Conder/Fauth/Furrer:1998, Anisimov/Bukhvalov/Rice:1999, Zhou/Goodenough/Dabrowski:2005, Mazin/Khomskii/Lengsdorf/etal:2007, Catalan:2008,Lau/Millis:2013} have suggested the likely existence of a complex interplay between octahedral rotations, transport, and magnetic properties.
To extract deeper insight into these interrelationships, we 
($i$) identify all active distortion modes in each composition, 
($ii$) determine the individual amplitudes for each modes, and 
($iii$) explore the statistical correlation between individual distortion-modes and the 
physical properties, specifically $T_{MI}$ and $T_{N}$.

\subsubsection{Structure Decomposition}

We follow the procedure outlined by Campbell \emph{et al.} \cite{Campbell/Stokes/Tanner/Hatch:2006} to decompose the $P2_{1}/c$ monoclinic crystal structure data, obtained from previously published diffraction studies, \cite{Alonso/Martinez/Casais/etal:2000, Medarde/Fernandez/Lacorre:2008, Garcia/Aranda/Alonso:2009, Munoz/Alonso/Martinez/Fernandez:2009} into the orthonormal symmetry-modes. 
Diffraction data for YNiO$_{3}$, ErNiO$_{3}$, LuNiO$_{3}$, and HoNiO$_{3}$ were measured at 295~K; DyNiO$_{3}$ at 200~K; NdNiO$_{3}$ at 50~K; and finally, PrNiO$_{3}$ at 10~K. $T_{MI}$ and $T_{N}$ were obtained from the review article by Catalan \cite{Catalan:2008}.

\begin{figure*}[t]
  \begin{center}
    \begin{tabular}{lrcc}
      (a) Undistorted octahedra & (b) $M_{2}^{+}$ $d$-type Jahn-Teller \\
       \resizebox{32mm}{!}{\includegraphics{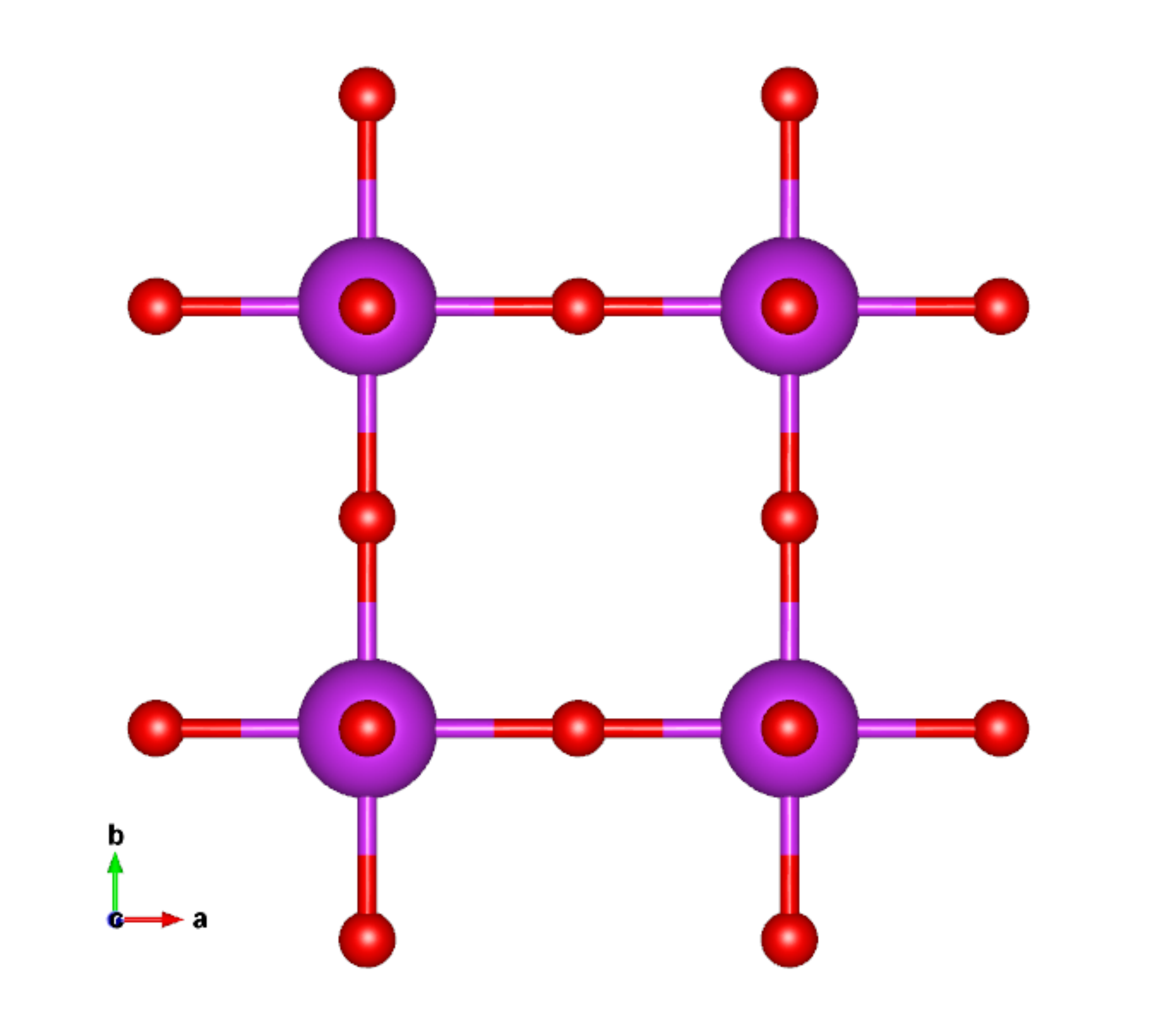}} &
       \resizebox{36mm}{!}{\includegraphics{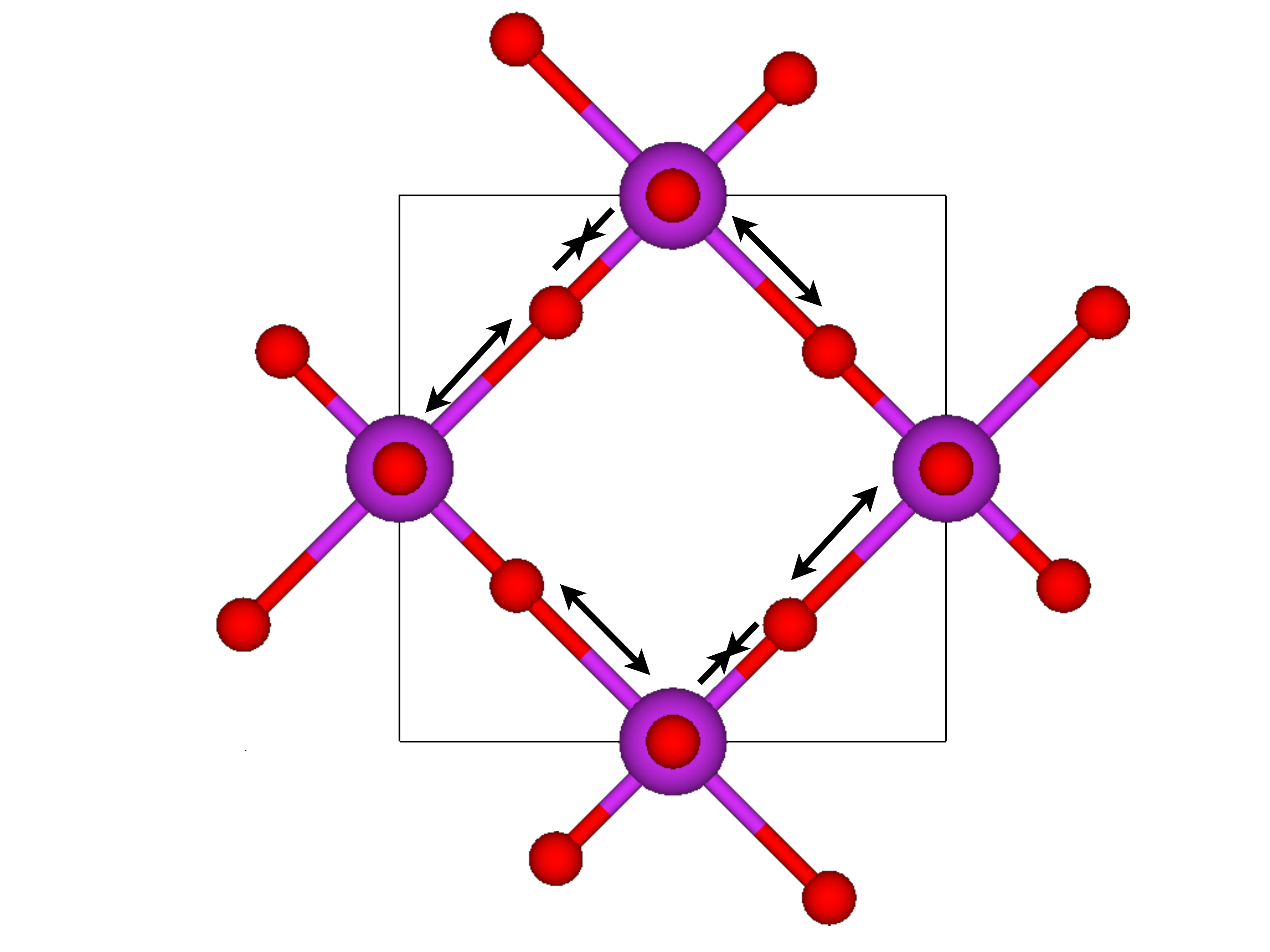}} \\
       (c) $M_{3}^{+}$ in-phase rotation & (d) $M_{5}^{+}$ out-of-phase tilting \\
       \resizebox{35mm}{!}{\includegraphics{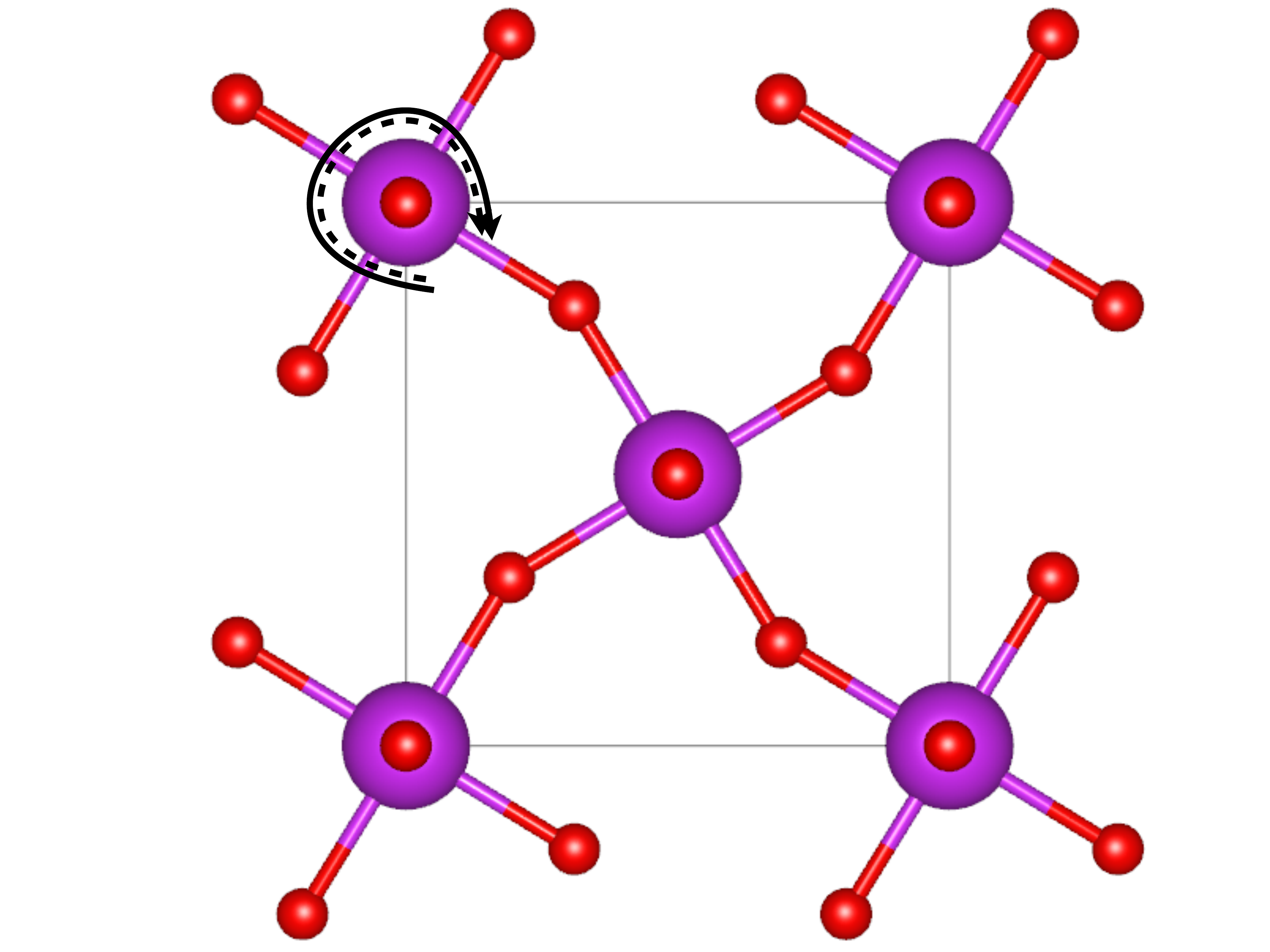}} &
       \resizebox{35mm}{!}{\includegraphics{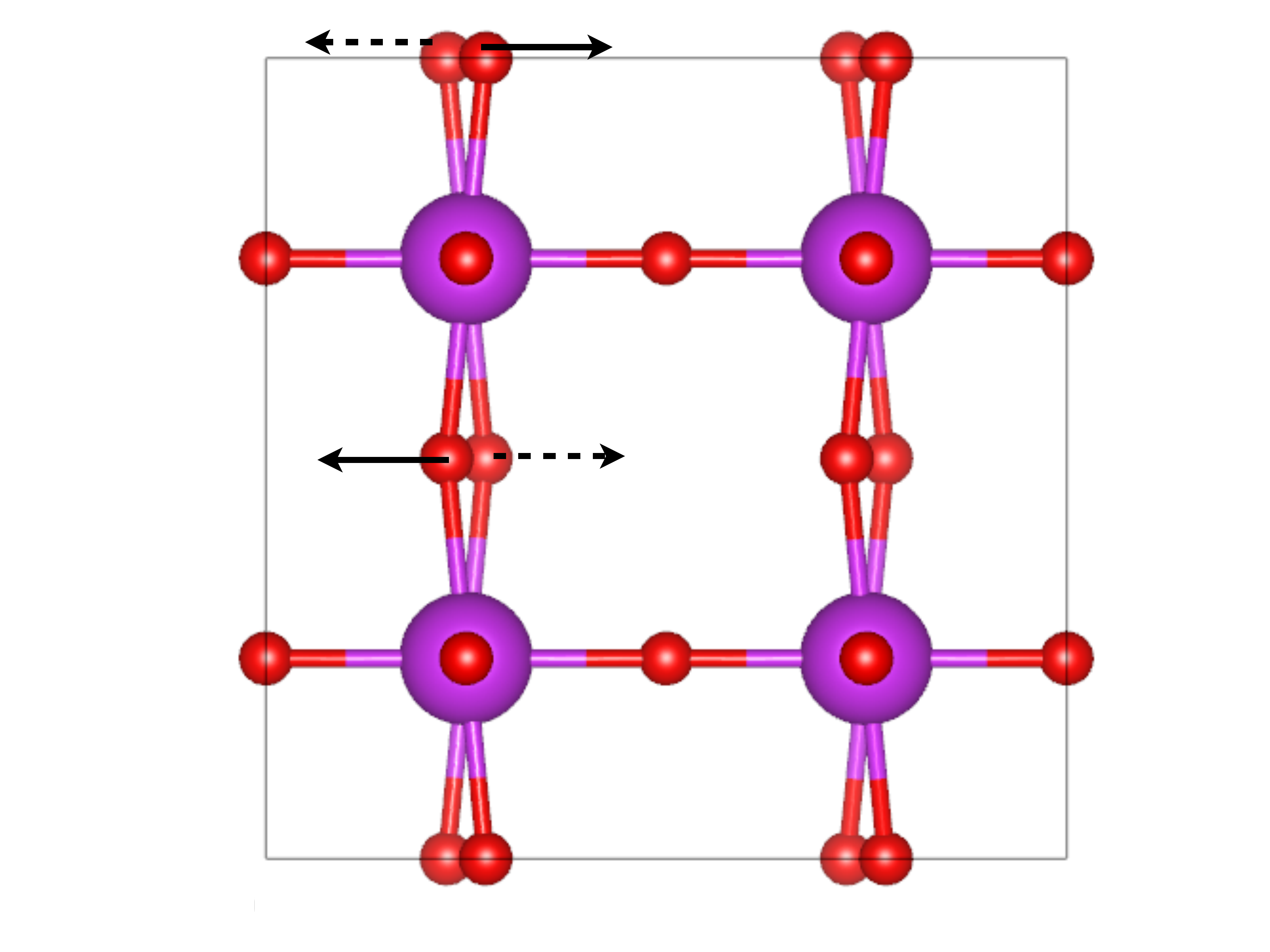}} \\
       (e) $R_{3}^{+}$ $a$-type Jahn-Teller  & (f) $R_{4}^{+}$ out-of-phase  rotation \\       
       \resizebox{35mm}{!}{\includegraphics{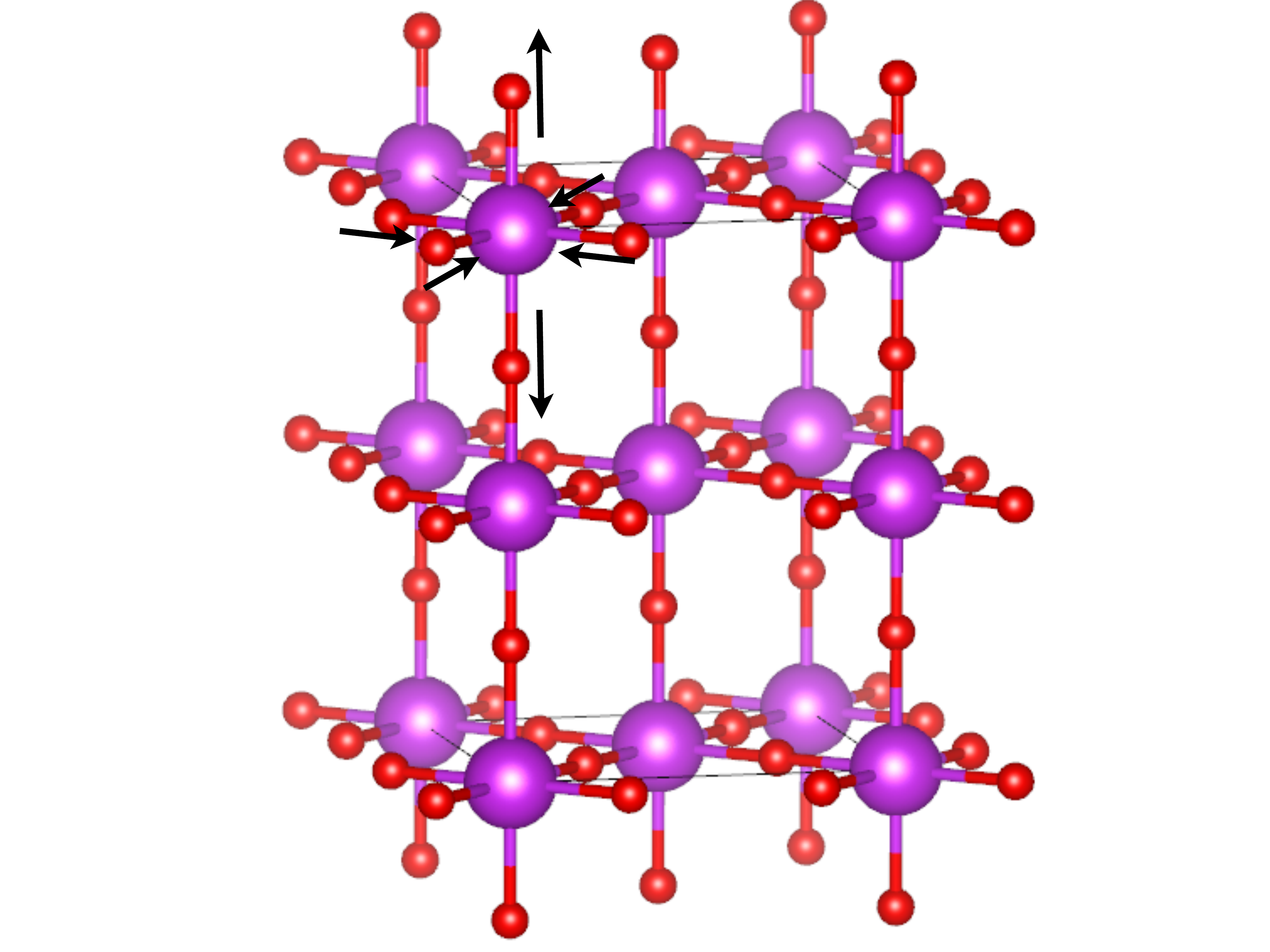}} &
       \resizebox{35mm}{!}{\includegraphics{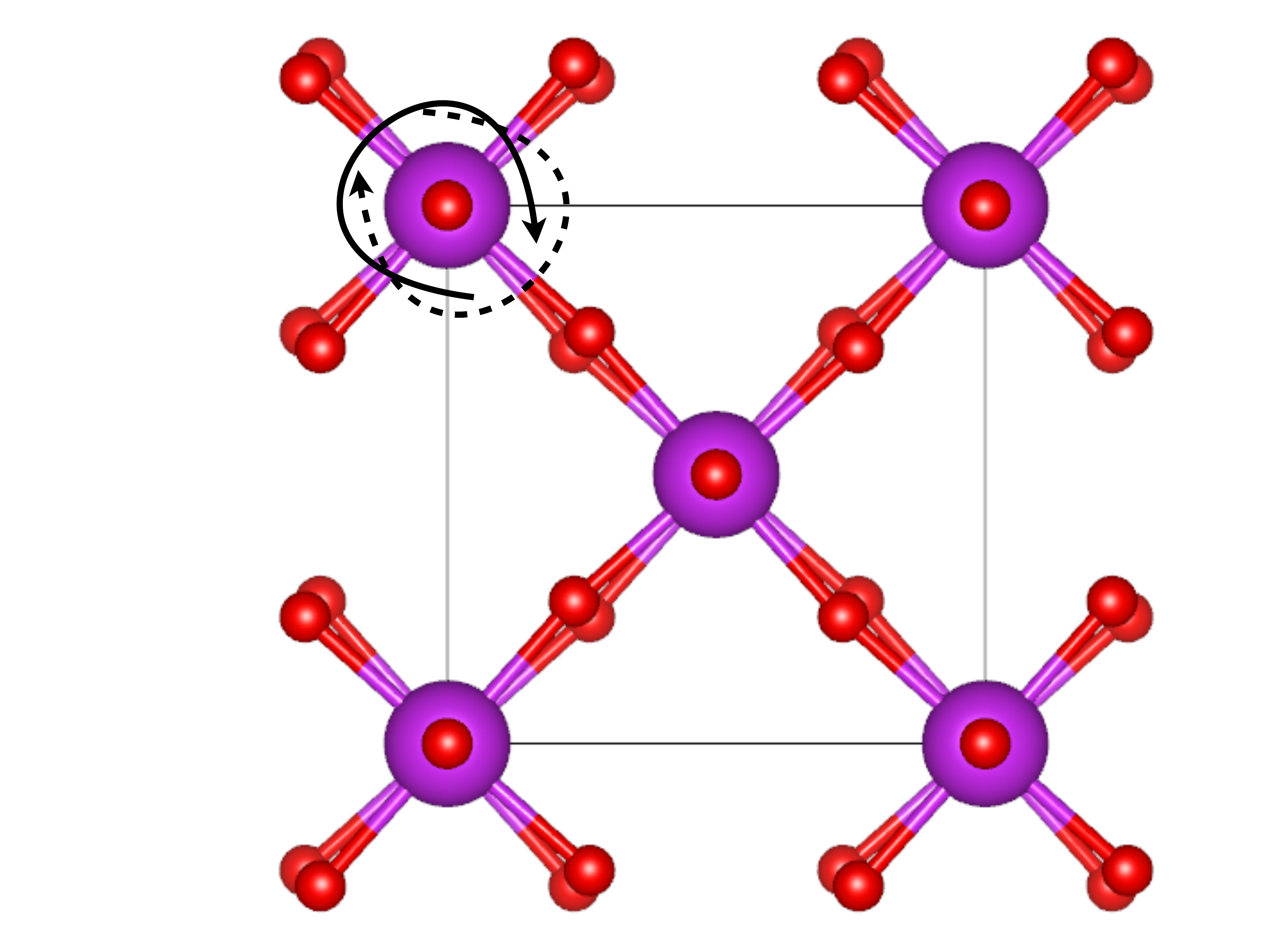}} \\
       (g) $R_{5}^{+}$ bending/buckling mode & (h) $X_{5}^{+}$ in-phase tilting\\              
       \resizebox{36mm}{!}{\includegraphics{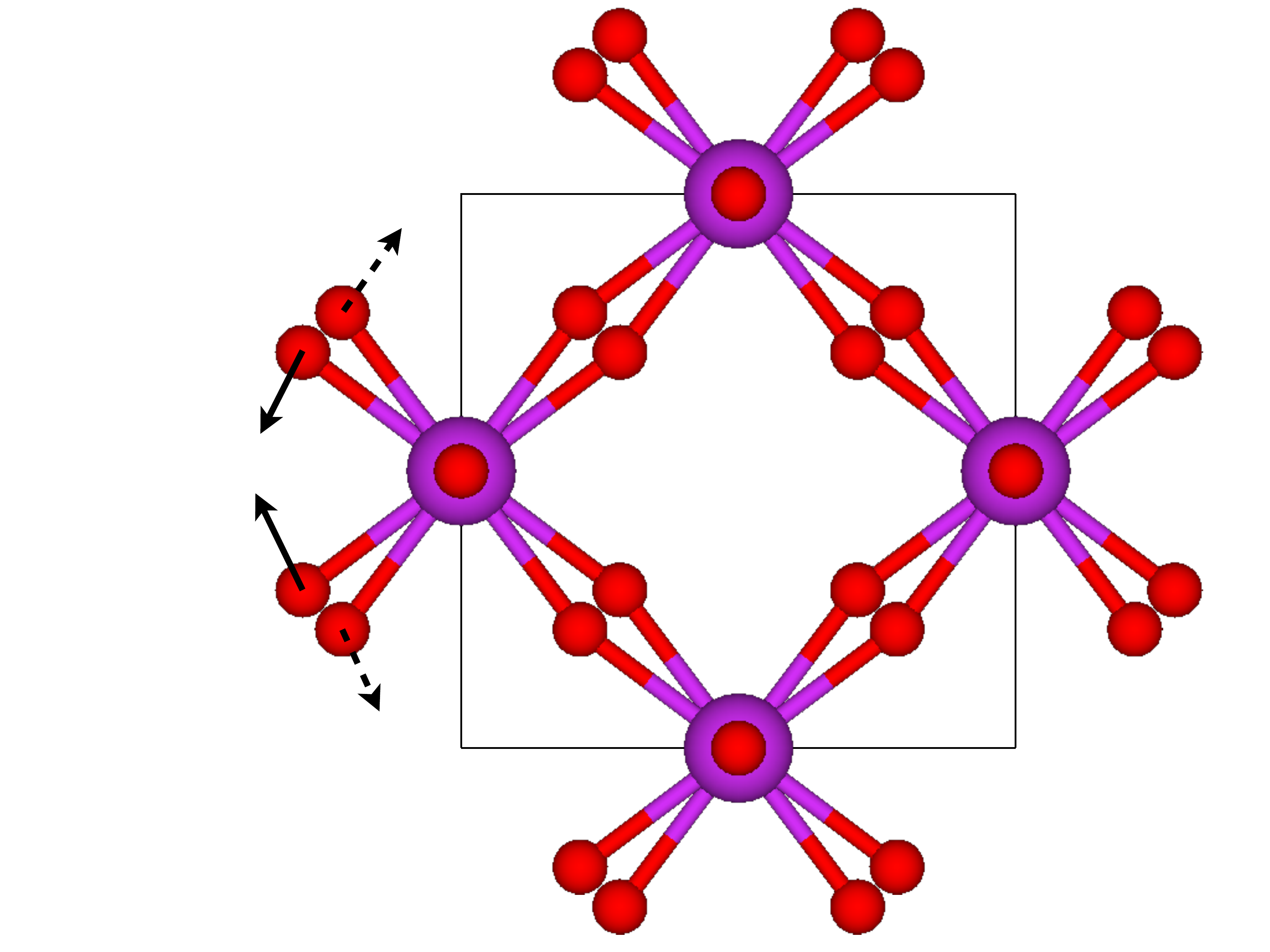}} &
       \resizebox{36mm}{!}{\includegraphics{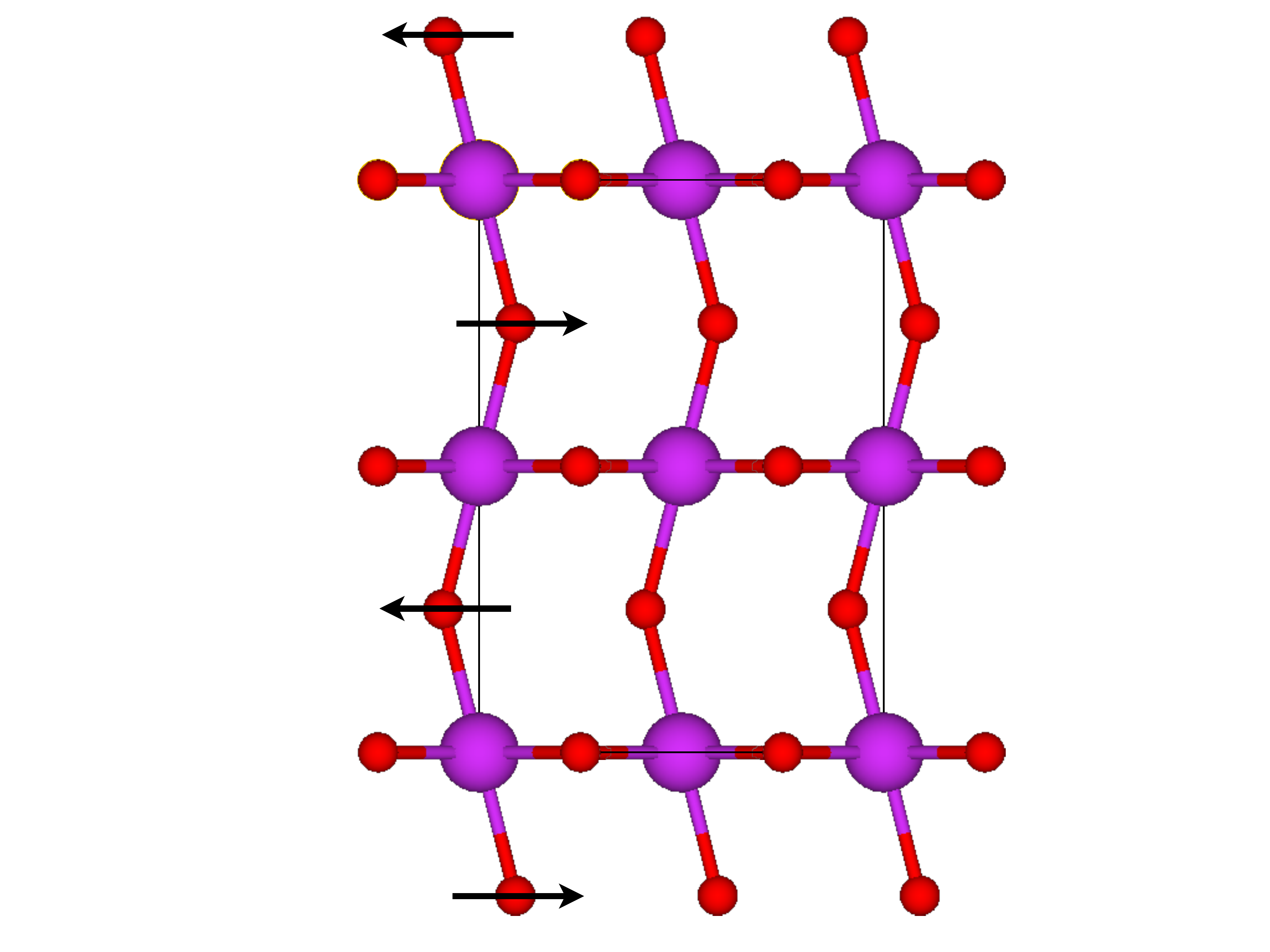}} 
    \end{tabular}\vspace{-5pt}
    \caption{(Color online.) Illustration of the 
	symmetry-adapted orthonormal distortion modes found in the low-symmetry 
	$R$NiO$_{3}$ and Ba$_{1-x}$K$_{x}$BiO$_{3}$ perovskites. 
	The undistorted octahedra are shown in panel (a) for comparison. Octahedral tilting distortions refer to the rotation of the octahedral units (we use `tilting' and `rotation' interchangeably). The out-of-phase distortions are differentiated using arrows (lines and dashes) indicating the direction of cooperative atomic displacements. 
(b) $M_{2}^{+}$ describes the $d$-type Jahn-Teller  mode, where two bonds shrink and two elongate; 
(c) $M_{3}^{+}$ describes an in-phase rotation mode; 
(d) $M_{5}^{+}$ describes an out-of-phase tilting mode, which could also act as an in-phase tilting mode depending on the order parameter direction 
(e) $R_{3}^{+}$ describes the $a$-type Jahn-Teller mode, where four bonds contract and two expand, and \emph{vice versa}; 
(f) $R_{4}^{+}$ describes an out-of-phase rotation mode; 
(g) $R_{5}^{+}$ describes an out-of-phase bending mode; 
(h) $X_{5}^{+}$ describes an in-phase tilting mode accompanied by $A$ cation displacements, which 
are not shown for clarity.}
    \label{fig:tilting}
  \end{center}
\end{figure*}

The $P2_{1}/c$ structure decomposes into eight algebraically independent 
symmetry-modes corresponding to the following irreps: 
$R_{1}^{+}$, $R_{3}^{+}$, $R_{4}^{+}$, $R_{5}^{+}$, $X_{5}^{+}$, $M_{2}^{+}$, $M_{3}^{+}$, and $M_{5}^{+}$. 
The relative amplitudes for each of the distortions are given in \autoref{tab:op}, with 
the physical representation of each irrep and its consequence on the octahedral 
framework schematically illustrated in \autoref{fig:tilting}. 

\begingroup
\squeezetable
\begin{table}[b]
\begin{ruledtabular}
\caption{Summary of the amplitudes of each irrep (notation given with respect to the $Pm\bar{3}m$ parent space-group) in the $P2_{1}/c$ structure (units in {\AA}). Std.\ Err.\ = $\sigma/N$, where $\sigma$ is the standard deviation and $N$ (=7) is the sample size, gives an estimate of the sampling error. The total distortion amplitude is given in the column indicated by $\sum$.}
\centering
\begin{tabular}{llllllllll}
	&	\multicolumn{5}{c}{also present in $Pnma$} & & & & \\ 
\cline{2-6}\\[-2.7ex]
$R$NiO$_{3}$	&$R_{4}^{+}$	&$R_{5}^{+}$	&$X_{5}^{+}$	&$M_{2}^{+}$	&$M_{3}^{+}$	&$R_{1}^{+}$	&$R_{3}^{+}$	&$M_{5}^{+}$& $\sum$	 \\[1.ex]
\hline
HoNiO$_{3}$	&1.499	&0.195	&0.826	&0.048	&1.153	&0.172	&0.024	&0.025 & 2.081\\
ErNiO$_{3}$	&1.535	&0.203	&0.842	&0.048	&1.179	&0.145	&0.022	&0.020 & 2.127\\
LuNiO$_{3}$	&1.648	&0.248	&0.907	&0.063	&1.229	&0.153	&0.004	&0.016 & 2.268\\
YNiO$_{3}$	&1.502	&0.196	&0.835	&0.055	&1.161	&0.128	&0.011	&0.072 & 2.089\\
NdNiO$_{3}$	&1.179	&0.117	&0.445	&0.030	&0.751	&0.126	&0.050	&0.283 & 1.505\\
PrNiO$_{3}$	&1.093	&0.063	&0.362	&0.004	&0.690	&0.091	&0.012	&0.000 & 1.347\\
DyNiO$_{3}$	&1.442	&0.207	&0.780	&0.073	&1.144	&0.241	&0.025	&0.168 & 2.041\\
\hline
Std.\ Err.\	&0.029	&0.009	&0.031	&0.003	& 0.032	&0.007	&0.002	&0.015& --	\\	
\end{tabular}
\label{tab:op}
\end{ruledtabular}
\end{table}
\endgroup

The sequence of structures involved in the phase transition could be written as follows: $Pm\bar{3}m$$\rightarrow$$Pnma$$\rightarrow$$P2_{1}/c$. 
Five distortions participate in the $Pm\bar{3}m$$\rightarrow$$Pnma$ transition, namely $R_{4}^{+}$, $R_{5}^{+}$, $X_{5}^{+}$, $M_{2}^{+}$, and $M_{3}^{+}$. 
Within Landau theory, the $Pnma$ structure results from the condensation of two  zone-boundary phonons of different wavevectors:
one at the $R$-point of the cubic Brillouin zone that transforms as the 
irrep $R_{4}^{+}$,  and the other at the $M$-point, which transforms as the irrep $M_{3}^{+}$. 
Not surprising, the two most common octahedral distortion-modes that describe the $a^+b^-b^-$ rotation pattern, $R_{4}^{+}$ and $M_{3}^{+}$, emerge as the  primary modes. A secondary mode with $X_{5}^{+}$ symmetry also appears with relatively high amplitude, and it contains $A$ cation displacements. 
Such anti-parallel displacements of the $A$ cations are established to  be correlated 
with the amplitudes of the octahedral rotations in perovskites.\cite{Mulder/Rondinelli/Fennie:2013}
In the symmetry lowering $Pnma$$\rightarrow$$P2_{1}/c$ structural phase transition, irrep $R_{1}^{+}$ with order parameter $(a)$ is the primary order parameter capturing the CDP behavior. The $R_{3}^{+}$ and $M_{5}^{+}$ are secondary distortions in the phase transition that accompany $R_{1}^{+}$.
This is seen in \autoref{tab:op}, where the largest amplitude for modes not already present from the transition to $Pnma$ is almost always given by $R_{1}^{+}$.
%
%
%
NdNiO$_{3}$ is found to be the exception to this rule and the origin of $Pnma$$\rightarrow$$P2_{1}/c$ phase transition in NdNiO$_{3}$ could be different compared to other bulk RNiO$_{3}$ compounds.

In the nickelate literature, generally NdNiO$_{3}$ is discussed in conjunction with PrNiO$_{3}$, because together they represent a unique case where 
$T_{MI}$ and $T_{N}$ coincide.
In NdNiO$_{3}$, we find that the distortion amplitude for $M_{5}^{+}$ is twice that of $R_{1}^{+}$, indicating that irrep $M_{5}^{+}$ acts as the primary distortion mode instead of $R_{1}^{+}$. 
Indeed our group theoretical analysis shows that for the direct sum, $M_{3}^{+}$$\oplus$$R_{4}^{+}$$\oplus$$M_{5}^{+}$, whose tilt system correspond to the order parameter containing only three free parameters:$(a,0,0,0,b,b,c,c,0,0,c,-c)$, the required space group is $P2_{1}/c$ monoclinic structure, an isotropy subgroup of $Pnma$, indicating that $M_{5}^{+}$ could be the primary distortion mode in the phase transition if it also provides the greatest energetic stability to the structure. The latter constraint requires evaluation of the lattice dynamics 
using a technique beyond group theory.
Interestingly for the experimental  bulk PrNiO$_{3}$ structure, irrep $M_{5}^{+}$ is found to have zero amplitude.
To summarize this discussion, we have introduced an alternative symmetry-mode 
description of the structural distortions in nickelates that identifies bulk NdNiO$_{3}$ compound as anomalous in the series. Combining temperature-dependent diffraction studies with mode decomposition analysis could provide  key insights necessary to fully understand the evolution of the primary and secondary distortions with the thermodynamic origin of phase transition.

We now shift our attention to the weak secondary distortion modes, particularly to the observation of two types of Jahn-Teller distortions to the NiO$_6$ units, $M_{2}^{+}$ and $R_{3}^{+}$ (\autoref{fig:tilting}). Although $M_{2}^{+}$ and $R_{3}^{+}$ are practically negligible and are the secondary distortion modes in the $Pm\bar{3}m$$\rightarrow$$Pnma$$\rightarrow$$P2_{1}/c$  transition, their presence could still have significant implications on the physical properties. The motivation for critically analyzing secondary distortion modes in nickelates comes from the study of \emph{improper} phase transitions,\cite{Dvorak:1971,Levanyuk:1974,Toledano:1987} \emph{e.g.}, where spontaneous electric polarization arises as a secondary effect accompanying complex non-polar distortions through anharmonic interactions\cite{Perez-Mato/Blaha:2008,Etxebarria/Perez-Mato:2010,Benedek/Fennie:2011,Stroppa:2013} in improper ferroelectrics. Similarly, we are interested in exploring the role of secondary order parameters on the electronic properties of nickelates.
Experimental observations based on giant oxygen O$^{16}$-O$^{18}$ isotope effect \cite{Medarde/Lacorre/Conder/Fauth/Furrer:1998} indicate the presence of dynamic Jahn-Teller polarons; however, the interpretation of orbital-ordering has lacked conclusive support from diffraction experiments \cite{Catalan:2008}. 
More recently, ultra-thin films of nickelates are being increasingly investigated for rational control of orbital polarization \cite{Chakhalian/Rondinelli/etal:2011, Freeland/Liu/etal:2011, Chen/Ismail-Beigi:2013}. The identification of a Jahn-Teller bond elongations and contractions from our symmetry analysis  in  the bulk compounds suggests the nickelates should have tendencies to  orbital (ordering) polarizations. Misfit strain, superlattice formation, or symmetry mismatch at the thin film--substrate interface may be used to selectively enhance the contributions of Jahn-Teller modes, described by irreps $M_{2}^{+}$ and $R_{3}^{+}$, as recently proposed in Ref.~\onlinecite{Freeland:Submitted}. Alternatively, orbital-ordering could \emph{emerge} in nickelates, \emph{e.g.}, LaNiO$_{3}$ where the Jahn-Teller modes are prohibited by symmetry in the bulk ground-state structure, through quantum confinement effects in the limits of ultra-thin films \cite{Chakhalian/Rondinelli/etal:2011, Freeland/Liu/etal:2011}; although in this case, strain-induced Jahn-Teller distortions would likely result in larger energetic penalties.

\subsubsection{Statistical Analysis\label{sec:statistical}}
We now evaluate the statistical correlation between the amplitude of the symmetry-adapted modes and the macroscopic $T_{MI}$, and $T_{N}$ transition temperatures. The physical motivation behind the statistical analysis is to uncover hidden associations between the cooperative atomic displacements and electronic/magnetic phase transitions. The goal is to quantitatively identify structure--functionality relationships that could be rigorously evaluated at the \emph{ab initio} level.

\begin{figure}
\centering
   \includegraphics[width=0.99\columnwidth,clip]{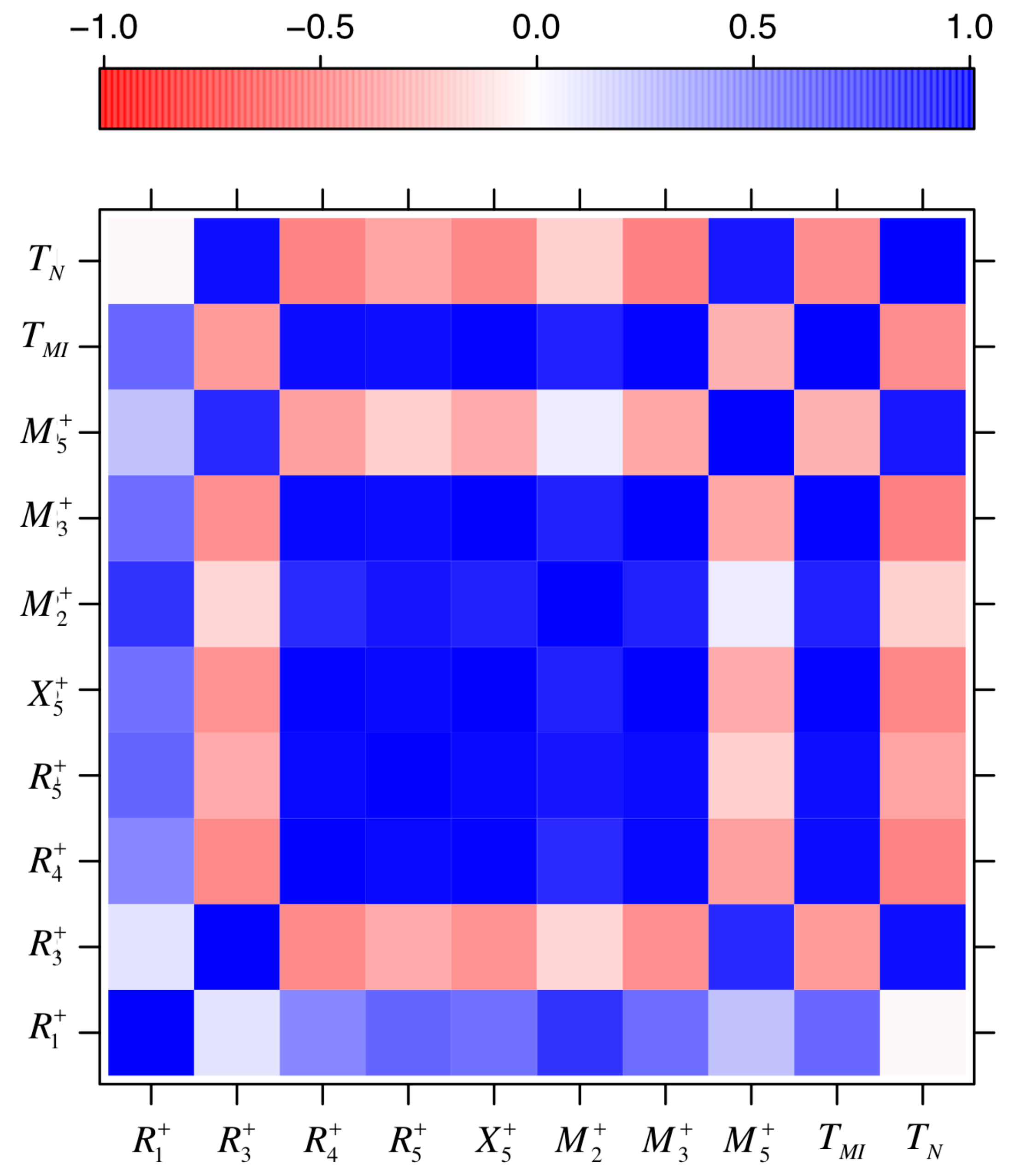}\vspace{-6pt}
    \caption{(Color online.) Correlation heat map capturing the degree of linear 
	relationship between unit cell level structural distortion modes and macroscopic 
	transition temperatures, $T_{MI}$, and $T_{N}$ in $R$NiO$_{3}$ perovskites.
	Dark blue (red) indicates a strong positive (negative) correlation and white 
	indicates no statistical correlation between between the two variables. 
	Units for the irreps and temperatures are given in {\AA} and Kelvin (K), respectively.}
    \label{fig:corr_map}
\end{figure}

We begin by constructing a dataset containing seven $R$NiO$_{3}$ nickelates, where each 
rare-earth nickelate in our set is described using eight distortion-modes, 
$T_{MI}$, and $T_{N}$, resulting in a $7\times10$ matrix.
%
%
We scaled the data by subtracting the mean of each column from its corresponding 
columns (this process is also known as centering), and then divided the 
centered columns by their standard deviations. 
To evaluate the degree of linear relationship between the structural distortions, 
$T_{MI}$, and $T_{N}$, we calculate the sample covariance of the centered and 
scaled dataset as 
\[
\mathrm{Cov}[X] = \frac{1}{n-1}\sum\limits_{i=1}^n (\bar{x}_i)(\bar{x}_j)\, ,
\]
where $n=7$ is the total number of $R$NiO$_{3}$ compounds in our dataset; $\bar{x}_{i}$ and $\bar{x}_{j}$ are the centered and scaled column vectors of our data matrix $X$, respectively. 
Results from the covariance analysis are summarized as a correlation heat map in \autoref{fig:corr_map}.

A strong positive correlation is found between irreps that describe distortions 
to the NiO$_6$ octahedra: $M_{2}^{+}$ 
(a planar Jahn-Teller mode), $M_{3}^{+}$, $X_{5}^{+}$, $R_{4}^{+}$, and 
$R_{5}^{+}$ modes. 
Even though the distortion-modes are orthonormal by construction, when they 
are collectively evaluated for a series of $R$NiO$_{3}$ compounds, our analysis 
reveals that they (the modes describing the distortions) are statistically \emph{dependent} and coupled. 
These modes largely describe bond angle distortions 
and are positively correlated with the electronic transition $T_{MI}$.
The conventional route to describe variations in $T_{MI}$ primarily focus on 
tolerance factor and \emph{average} $\langle$Ni--O--Ni$\rangle$ bond angle, 
whereby bending of the $\langle$Ni--O--Ni$\rangle$ angle further from the ideal case of 180$^\circ$ 
decreases the bandwidth, promoting the insulating state over the metallic state\cite{Torrance/Lacorre/Nazzal/etal:1992,Obradors/Paulius/Maple/etal:1993, Catalan:2008}.
While we recover this behavior,  we also identify the unique 
displacement patterns that geometrically sum to give the aggregate bond angle: 
irreps $M_{2}^{+}$, $M_{3}^{+}$, $X_{5}^{+}$, $R_{4}^{+}$, and $R_{5}^{+}$ cooperatively act to bend the Ni--O--Ni angle.
These five irreps fully describe the $Pnma$ crystal structure relative to the 
cubic phase  found in the metallic nickelates 
at high-temperature, reinforcing the concept that the  
orthorhombic distortions are largely responsible for the bandwidth-controlled transport behavior in nickelates, and hence \emph{prepare} the electronic system for the MIT.
Intriguingly, the $R_{1}^{+}$ CBD distortion, which is the usual signature for CO has only a
moderate effect on $T_{MI}$, indicating that the dominant structural route to 
engineer the electronic transition may not be solely through \emph{isotropic} bond length 
distortions.
Moreover, \emph{anisotropic} bond distortions obtained with the $a$-type $R_{3}^{+}$ Jahn-Teller mode 
or the $M_{5}^{+}$ tilting mode do not contribute significantly to $T_{MI}$. In fact, we find they are  
\emph{anti-correlated} with the  electronic transition temperature (\autoref{fig:corr_map}). 
Our conclusion to this point is that the geometry (tilt pattern) of the oxygen framework structure 
is the important atomic scale feature governing the MI-transition.

We now shift our attention to the Type-E$^\prime$ antiferromagnetic ordering. 
Unlike the transport behavior, where the $\langle$Ni--O--Ni$\rangle$ bond angle concept 
appears to be sufficient to explain its variability, the origin of antiferromagnetic ordering 
is much more complex. 
Previous high-resolution photoemission measurement \cite{Vobornik/etal:1999} and 
pressure dependent studies \cite{Zhou/Goodenough/Dabrowski:2005} of $T_{N}$ 
have established the existence of two distinct regimes for $R$NiO$_{3}$ with 
$T_{MI}=T_{N}$ and $T_{MI}>T_{N}$, 
indicating that simple tolerance factor and $\langle$Ni--O--Ni$\rangle$ superexchange interaction 
arguments are insufficient.
%
Lee \emph{et al.} used two-band model and Hartree-Fock theory to classify 
nickelates with $T_{MI}=T_{N}$ as itinerant, whereas those with $T_{MI}>T_{N}$ 
show stronger electron correlation.\cite{Lee:2011, Lau/Millis:2013}  
Nonetheless, we are unaware of a unique descriptor available in the literature 
that captures all the variance in the  $T_{N}$ ordering temperature.

\begin{figure}
\centering
   \includegraphics[width=0.99\columnwidth]{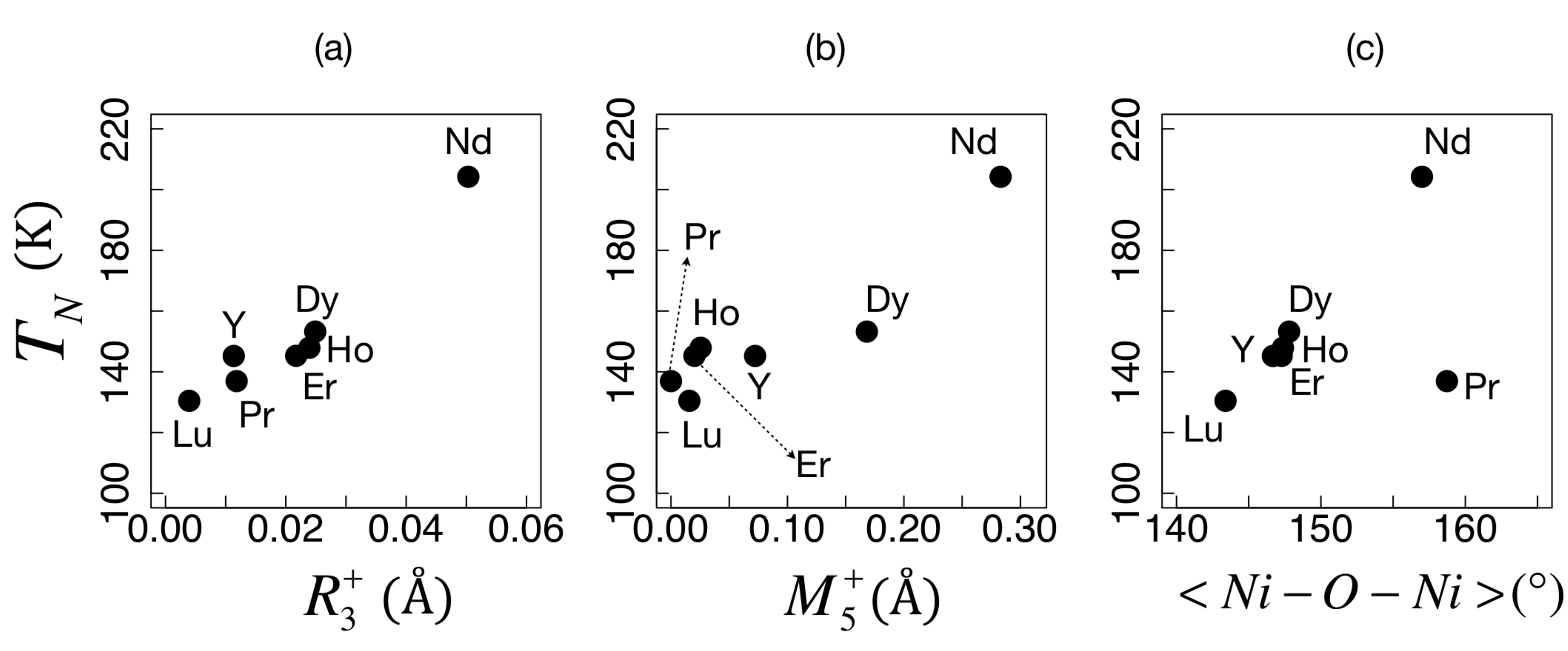}\vspace{-10pt}
    \caption{(Color online.) A strong linear relationship is seen between (a) $R_{3}^{+}$ and the N\'{e}el temperature ($T_{N}$) with a correlation coefficient of 0.957 and (b) $M_{5}^{+}$ and $T_{N}$ with a correlation coefficient of 0.916. In (c) we plot the average $\langle$Ni--O--Ni$\rangle$ bond angle and $T_{N}$  to illustrate that the  $R_{3}^{+}$ and $M_{5}^{+}$ irreps  capture the main variation in $T_{N}$ for both $T_{MI}=T_{N}$ and $T_{MI}>T_{N}$ nickelates compared to the $\langle$Ni--O--Ni$\rangle$ bond angle. Experimental data from Ref.~\onlinecite{Catalan:2008}.
}
\label{fig:neel}
\end{figure}

Our analysis reveals the existence of a strong linear relationship between $T_{N}$ and two irreps, 
$R_{3}^{+}$ and $M_{5}^{+}$: the three dimensional Jahn-Teller mode, 
where in each NiO$_6$ octahedron four bonds contract and two elongate, 
and the in-phase tilting mode, respectively.
The linear relationship is valid for both $T_{MI}=T_{N}$ and $T_{MI}>T_{N}$ 
nickelates (\autoref{fig:neel}), indicating that $R_3^+$ \emph{and}  
$M_5^+$ contain additional information that is not captured by 
either the conventional $\langle$Ni--O--Ni$\rangle$ angle or tolerance factor descriptors. 

The $R_{3}^{+}$ distortion is more strongly correlated with $T_N$ 
than the  $M_{5}^{+}$, indicating an underlying structural relationship between the 
long-range antiferromagnetic order and the local and subtle Jahn-Teller bond distortions. 
We also note that $R_{3}^{+}$ and $M_{5}^{+}$ irreps appear in the 
monoclinic $P2_{1}/c$ crystal structure and are not allowed by symmetry in the $Pnma$  
orthorhombic space group.
New insights gained thus far support the potential feasibility of tailoring an isosymmetric $P2_{1}/c \rightarrow P2_{1}/c$ phase transition driven by misfit strain, interfaces and surfaces, thermodynamic state variables, and/or any combinations thereof, which would result in the preferential enhancement of atomic displacements transforming as $R_{3}^{+}$ and $M_{5}^{+}$ distortions relative to $R_{1}^{+}$. 

We conjecture that with deterministic control over the atomic 
structure, \emph{e.g.}, enhancing $R_{3}^{+}$ relative to other distortion modes, it could be possible to fully decouple $T_{MI}$ from $T_N$, shifting the equilibrium phase boundaries dividing the metallic/insulating states from the paramagnetic/ordered regions to create non-equilibrium phases. 
Although it is compelling to extend our interpretation beyond correlation to full causation, 
we emphasize that first-principles calculations and detailed structural characterization 
are essential to establish the physical origin underlying the statistical associations we have 
identified.

\subsection{{{Ba}}$_{1-x}${{K}}$_{x}${{BiO}}$_{3}$ Bismuthates}

BaBiO$_{3}$ is a CO insulator which undergoes a series of 
phase transitions with temperature.
At room temperature it has a monoclinic crystal structure with space 
group $I2/m$. 
In the monoclinic $I2/m$ structure, the octahedrally coordinated tetravalent Bi cation 
charge disproportionates into Bi$^{(4+\delta)+}$ and Bi$^{(4-\delta)+}$. 
Previous optical conductivity measurements\cite{Karlow/etal:1993} attribute 
the appearance of an electronic gap in the stoichiometric $I2/m$ phase to 
structural CBD and octahedra rotations.

Chemical doping of BaBiO$_{3}$ with potassium by random substitution on 
the Ba sites, Ba$_{1-x}$K$_{x}$BiO$_{3}$, produces a sequence of structural 
and electronic phase transitions.
\emph{Increasing} the potassium concentration leads to crystal structures 
of \emph{higher} symmetry: 
An orthorhombic $Ibmm$ phase is stable over the range $0.12\le x <  0.37$.
At $x\approx0.37$, a tetragonal $I4/mcm$ phase is stable and for $0.37<x<0.53$ BaBiO$_{3}$ transforms to $Pm\bar{3}m$ cubic \cite{Braden/etal:2000, Franchini/Krese/Podloucky:2009}. 
The electronic transport properties also evolve concomitantly with these structural changes: The system  transforms from a robust insulating state in the monoclinic phase to semiconducting in the orthorhombic phase, and finally to a superconducting state in the tetragonal and undistorted cubic phases.
Although there are strong evidences supporting an apparent association between structure, doping, and onset of superconductivity, a quantitative assessment of the relationship linking structure to superconductivity is still lacking.
Our objective here is to decompose the crystal structure of Ba$_{1-x}$K$_{x}$BiO$_{3}$ as a function of K-doping for $0 \le x \le 0.4$ with emphasis placed on the role of the symmetry-adapted modes and CBD--rotation symmetries.

\begin{figure*}
\centering
   \includegraphics[width=1.45\columnwidth,clip]{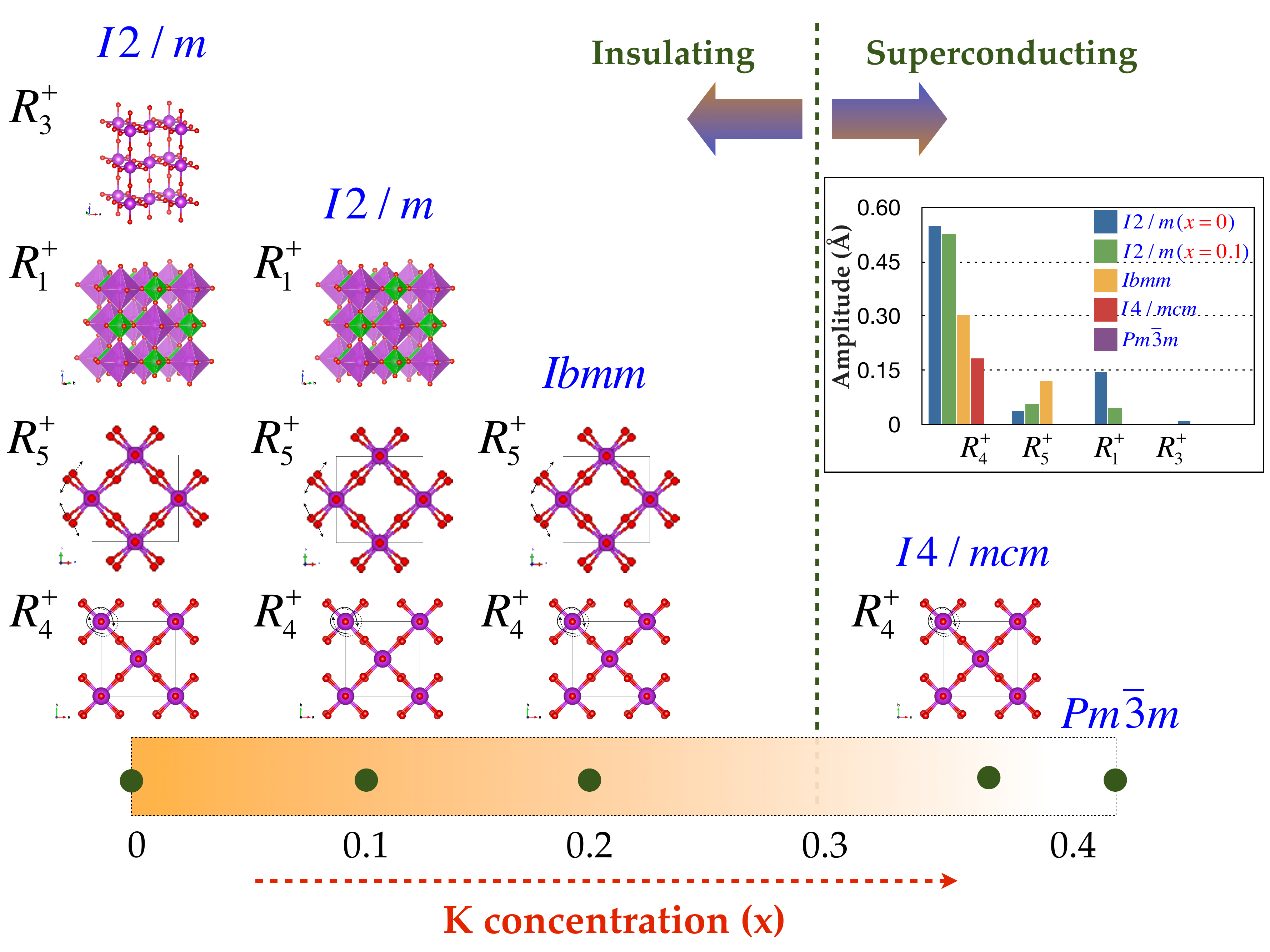}\vspace{-8pt}
    \caption{(Color online.) Relationship between crystal symmetry and structural 
	distortions in Ba$_{1-x}$K$_{x}$BiO$_{3}$ as a function of $A$ site doping. 
	The number of symmetry unique structural modes decreases with 
	increasing K concentration. 
	Pure BaBiO$_{3}$ is described by four modes, $R_{1}^{+}$, $R_{3}^{+}$, 
	$R_{4}^{+}$, and $R_{5}^{+}$. 
	As the concentration of K increases, the Jahn-Teller $R_{3}^{+}$ mode 
	disappears rapidly, followed by loss of the three-dimensional CBD $R_{1}^{+}$.
	The bar graph (inset) shows quantitatively in units of angstroms the change in amplitude 
	of the structural modes observed in each space group.
}
    \label{fig:babio3_modes}
\end{figure*}

We decompose the low-symmetry crystal structures of Ba$_{1-x}$K$_{x}$BiO$_{3}$ 
across the phase boundaries using five chemical compositions: 
BaBiO$_{3}$,  
Ba$_{0.9}$K$_{0.1}$BiO$_{3}$, 
Ba$_{0.8}$K$_{0.2}$BiO$_{3}$, 
Ba$_{0.63}$K$_{0.37}$BiO$_{3}$, and 
Ba$_{0.60}$K$_{0.40}$BiO$_{3}$, 
for which high-resolution diffraction data exists.\cite{Pei/etal:1990,Pei/etal:1990,Pei/etal:1990,Braden/etal:2000,Pei/etal:1990}
When we constructed the hypothetical parent reference structure, we assigned to both K and Ba atoms the same $(x,y,z)$ coordinates and we employed fractional site occupancies to treat the random K-substitution on the Ba-site. This treatment ensures that the parent structure is $Pm\bar{3}m$ as opposed to any other space group, in which case the $(x,y,z)$ coordinates for the two cations would be different in the unit cell, \emph{i.e.}, the $A$-site Wyckoff orbit would split into two, and the space group of the resulting structure would not be the cubic $Pm\bar{3}m$ aristotype. 
%

In practice, the occupancy mode amplitudes are calculated by taking into account the site multiplicity and the fractional occupancy. In this work, for consistency, the fractional site occupancies for the Ba and K atoms in both the hypothetical parent and the experimentally observed distorted structures were kept the same. 
As a result, 
$Pm\bar{3}m$ (or for that matter $I2/m$, $Ibmm$, and $I4/mcm$) are treated as random solid solutions, rather than with periodic ordering.
Although the static symmetry modes are unable to accurately capture all the nuances of the configurational entropy governing random occupations of Ba/K-atoms, the amplitudes of the primary and secondary distortion modes do provide insight into the origin of the doping-dependent structural distortions reported in average structure x-ray diffraction measurements.
This approach is particularly useful to build an understanding of the interplay of doping and structure in the bismuthates, because computational modeling of Ba$_{1-x}$K$_{x}$BiO$_{3}$ phases present non-trivial challenges due to the random nature of the solid solution.
Such configurations necessitate the construction of large supercells and may require computationally-demanding hybrid functionals to accurately reproduce the electronic structure \cite{Franchini/Krese/Podloucky:2009}.

The results from our analysis are summarized in 
\autoref{fig:babio3_modes}. 
The $I2/m$ structure is described by four irreps 
$R_{1}^{+}$, $R_{3}^{+}$, $R_{4}^{+}$, and $R_{5}^{+}$, 
which capture the CBD, a Jahn-Teller distortion,  
out-of-phase octahedral rotations, and out-of-phase bond stretching, respectively.
Although the $I2/m$ monoclinic symmetry is maintained at small doping, as with 
Ba$_{0.9}$K$_{0.1}$BiO$_{3}$, our mode decomposition analysis reveals that
the amplitude of the $R_{3}^{+}$ distortion vanishes completely. 
The relative amplitudes of the $R_{1}^{+}$ CBD and out-of-phase 
octahedral rotation modes also decrease, whereas the amplitude of 
the out-of-phase bending distortion increases [\autoref{fig:babio3_modes} (inset)]. 
With further increase in K concentration, for example in Ba$_{0.8}$K$_{0.2}$BiO$_{3}$, 
the crystal structure becomes orthorhombic $Ibmm$. 
Interestingly, even in the absence of the CBD distortion ($R_1^+$), which we find has 
zero amplitude at $x=0.2$, the electronic structure remains insulating \cite{Franchini/Krese/Podloucky:2009}.
At the same time, the amplitude of the $R_{5}^{+}$ distortion 
continues to increase in the presence of out-of-phase octahedral rotations, described by $R_4^+$.
Upon further increase in K, the amplitude of $R_{5}^{+}$ distortion disappears completely, yielding 
tetragonal $I4/mcm$ Ba$_{0.63}$K$_{0.37}$BiO$_{3}$ with 
only the $R_{4}^{+}$ irrep, which correspond to modes with out-of-phase octahedral tilting at reduced amplitude.
Across this structural transition, the compound becomes metallic and 
superconducting\cite{Braden/etal:2000} with a critical transition temperature of 
27~K.
Finally, for $x\!\sim\!0.4$, 
the superconducting phase is also stable in the cubic $Pm\bar{3}m$ structure 
without any octahedral rotations or distortions.

An outstanding question concerning the electronic phase diagram of Ba$_{1-x}$K$_{x}$BiO$_{3}$ is the presence of the wide semiconducting or insulating region with doping. 
Our mode decomposition analysis provides an alternative interpretation of the phase diagram. 
In Ba$_{1-x}$K$_{x}$BiO$_{3}$, the octahedral rotations corresponding to irrep $R_{4}^{+}$ is found to be the primary distortion and it couples with other secondary distortions, namely $R_{3}^{+}$, $R_{1}^{+}$, and $R_{5}^{+}$ depending on the concentration of K. Particularly, we focus on the $R_{4}^{+}$ and $R_{5}^{+}$ distortions. Even though $R_{5}^{+}$ is a secondary distortion, it has a significant non-zero amplitude (0.12 {\AA}) near the insulator--superconductor phase boundary. To the best of our knowledge, no previous study has examined the role of primary and secondary distortions on the insulator--superconductor phase boundary.

In the insulating regime, the amplitude of $R_{4}^{+}$ monotonically decreases with increasing K. On the other hand, the amplitude of $R_{5}^{+}$ distortion is found to monotonically increase with K. The trend in $R_{4}^{+}$ and $R_{5}^{+}$ distortion pattern bears a remarkable resemblance with the evolution in the dielectric response reported by Nishio \emph{et al.} 
on  single crystal samples.\cite{Nishio/Ahmad/Hiromoto:2005} 
Nishio \emph{et al} identified two features in their optical spectra: the first was associated with the excitation from the lower Peierls band to the bipolaron defect state, which showed an increase in intensity with K-concentration similar to the increase in amplitude of $R_5^+$. 
The second feature was assigned with the magnitude of the Peierls gap, which decreased with increasing K concentration like $R_4^+$.
Our analysis suggests that the stability of the polaronic defects and the magnitude of the gap in the orthorhombic bismuthates may be intimately tied to the amplitude of the $R_{4}^{+}$ (rotations) and $R_{5}^{+}$ (bond stretching) distortions. 

Electronic structure calculations would allow us to unravel this association and glean 
insights into the persistence of wide semiconducting or insulating regions in Ba$_{1-x}$K$_{x}$BiO$_{3}$ compounds.
To do so, we propose a computational experiment to test the relationship between octahedral distortions and the optical spectra. One could ``freeze-in'' the $R_{4}^{+}$ (or alternatively $R_{5}^{+}$) distortion and map the energy landscape as a function of the amplitudes of $R_{5}^{+}$ (or the corresponding $R_{4}^{+}$) distortion. For each structures, the complex frequency-dependent dielectric function  would be computed following the computational approach of Franchini \emph{et al}, who has  successfully demonstrated that hybrid functionals are capable of describing both octahedral distortions and the complex electronic states.\cite{Franchini/Krese/Podloucky:2009} This would permit the independent assessment of  the relative role of $R_{4}^{+}$ and $R_{5}^{+}$ distortions on the closing of the Peierls gap and the stability of the bipolaron defect states. While our computational proposal may appear seemingly simple, this is by no means a trivial set of computations.

\section{Summary}
We applied group theoretical methods to evaluate the interplay between octahedral 
rotations and electronic charge ordering, which we parameterized structurally as 
octahedral ``breathing'' or cooperative bond length distortions through irreducible 
representations of the aristotype space group.
We enumerated the possible space groups available to perovskites from the 
intersection of these distortions.
This crystallographic data should prove useful for experimental 
structure refinements or serve as a well-defined set of symmetry-unique 
structures for which the stability of transition metal compounds 
susceptible to charge ordering may be evaluated with first-principles total energy methods.
With this information, we combined structural mode decomposition techniques 
with statistical methods to extract unbiased structure--property relationships from 
available experimental diffraction data on representative charge ordering oxides.
The crystal structures of perovskite nickelates and bismuthates were decomposed in terms 
of symmetry-adapted distortion-modes and the evolution in the amplitude of each 
mode tracked as a function of an external chemical parameter.
We illustrated that  this alternative set of descriptors provides a useful 
construct beyond the traditional tolerance factor paradigm found in perovskites 
to understand the atomic scale origin of physical properties, specifically how 
unit cell level modifications correlate with macroscopic functionality. 
Our statistical analysis uncovered previously unappreciated relationships that 
may be harnessed for electronic structure by design, 
some of which are being carried out within our group and will be reported later. 
We emphasize that the application of statistics to the decompositions does not 
require solely experimental data; the analysis may be performed with 
computationally obtained data alone.
Importantly, the relationships established with these techniques may be 
cross-validated by the construction of hybrid data sets, which combine 
theoretical results with experiment data, making it possible to extract and validate 
new insight into the material physics of oxides with correlated electrons.
We anticipate that this approach will spawn a number of additional studies in diverse
crystal classes since it is immediately generalizable: the synergy of applied 
group theoretical methods with statistical analysis and subsequent first-principles 
calculations provides a platform to achieve rational structure-driven 
design of complex materials.

\begin{acknowledgements}
P.V.B.\ was supported by The Defense Advanced Research Projects Agency (DARPA) under grant no.\ N66001-12-4224. J.M.R.\ was supported by the U.S.\ Office of Naval Research (ONR), under grant number N00014-11-1-0664. The views, opinions, and/or findings reported here are solely those of the authors and do not represent official views of DARPA or ONR. We thank our group members, D.\ Puggioni and A.\ Cammarata, for useful discussions.
\end{acknowledgements}

%

%
\bibliography{cbd_rot_paper_with_error_included_jmr}

\end{document}